\documentclass[twocolumn]{aastex62}

\usepackage{graphicx}
\usepackage{apjfonts}
\usepackage{amsmath}
\usepackage[raggedright]{subfigure}
\usepackage{color}
\usepackage{hyperref}
\usepackage{acronym}

\newcommand{\psr}{PSR\,J0952$-$0607}
\newcommand{\gamsrc}{4FGL\,J0952.1$-$0607}
\newcommand{\gcpsr}{PSR\,J1748$-$2446ad}
\newcommand{\rbpsr}{PSR\,J2339$-$0533}
\newcommand{\origpsr}{PSR\,B1937$+$21}
\newcommand{\fbwpsr}{PSR\,B1957$+$20}
\newcommand{\lowBpsr}{PSR\,J1544$+$4937}
\newcommand{\oLOFpsr}{PSR\,J1552$+$5437}
\newcommand{\blbwpsr}{PSR\,J1311$-$3430}

\def\Fermi{\textit{Fermi}}
\newcommand{\atlas}{ATLAS}

\def\DataDaysH{183 days}
\def\DataDaysL{169 days}
\def\DataDaysTot{707 days}
\def\hUL{h_0^{95\%}}
\def\epUL{\epsilon^{95\%}}
\def\hSD{h_0^{\mathrm{sd}}}
\def\epSD{\epsilon^{\mathrm{sd}}}
\def\twoF{2\mathcal{F}}
\newcommand{\hULval}{6.6 \times 10^{-26}}
\newcommand{\epULval}{3.1\times 10^{-8}}
\newcommand{\hSDval}{1.5 \times 10^{-27}}
\newcommand{\epSDval}{7.0 \times 10^{-10}}
\newcommand{\twoFcand}{9.9}

\newcommand{\distance}{1\,\mathrm{kpc}}
\newcommand{\calberr}{\pm 14\%}
\newcommand{\aboveSpindownLimitFactor}{45}

\newcommand{\pvec}{\boldsymbol{\lambda}}
\newcommand{\phiorb}{\phi_\text{orb}}
\newcommand{\tasc}{t_\text{asc}}
\newcommand{\tobs}{T_\text{obs}}
\newcommand{\fdot}{\dot{f}}
\newcommand{\Pdot}{\dot{P}}
\newcommand{\Bsurf}{B_\text{surf}}

\newcommand{\Porb}{P_\text{orb}}

\defcitealias{NE2001}{NE2001}
\defcitealias{YMW16}{YMW16}

\shorttitle{Gamma-Ray Detection of \psr} 
\shortauthors{\sc Nieder et al.}

\begin{document}
	
	\newacro{LAT}[LAT]{Large Area Telescope}
	\newacro{3FGL}[3FGL]{\Fermi~\acs{LAT} Third Source Catalog}
	\newacro{LOFAR}[LOFAR]{Low-Frequency Array}
	\newacro{GMRT}[GMRT]{Giant Metrewave Radio Telescope}
	\newacro{WFC}[WFC]{Wide Field Camera}
	\newacro{BIC}[BIC]{Bayesian Information Criterion}
	\newacro{HBA}[HBA]{high-band antenna}
	\newacro{MSP}[MSP]{millisecond pulsar}
	\newacro{SNR}[S/N]{signal-to-noise ratio}
	\newacro{IRF}[IRF]{instrument response function}

	\title{Detection and timing of gamma-ray pulsations from the $707$ Hz pulsar J0952$-$0607}

	\author[0000-0002-5775-8977]{L.~Nieder}
	\affiliation{Max-Planck-Institut f\"ur Gravitationsphysik (Albert-Einstein-Institut), 30167 Hannover, Germany}
	\affiliation{Leibniz Universit\"at Hannover, 30167 Hannover, Germany}
	
	\author[0000-0003-4355-3572]{C.~J.~Clark}
	\affiliation{Jodrell Bank Centre for Astrophysics, School of Physics and Astronomy, The University of Manchester, M13 9PL, UK}
	
	\author[0000-0002-1429-9010]{C.~G.~Bassa}
	\affiliation{ASTRON, The Netherlands Institute for Radio Astronomy, Oude Hoogeveensedijk 4, 7991 PD Dwingeloo, The Netherlands}
	
	\author[0000-0003-3536-4368]{J.~Wu}
	\affiliation{Max-Planck-Institut f\"ur Radioastronomie, Auf dem H\"ugel 69, 53121 Bonn, Germany}
	
	\author{A.~Singh}
	\affiliation{Max-Planck-Institut f\"ur Gravitationsphysik (Albert-Einstein-Institut), 30167 Hannover, Germany}
	\affiliation{Leibniz Universit\"at Hannover, 30167 Hannover, Germany}
	\affiliation{The Geophysical Institute, Bjerknes Centre for Climate Research, University of Bergen, Bergen 5007, Norway}
	
	\author{J.~Y.~Donner}
	\affiliation{Max-Planck-Institut f\"ur Radioastronomie, Auf dem H\"ugel 69, 53121 Bonn, Germany}
	\affiliation{Fakult\"at f\"ur Physik, Universit\"at Bielefeld, Postfach 100131, 33501 Bielefeld, Germany}
	
	\author[0000-0003-4285-6256]{B.~Allen}
	\affiliation{Max-Planck-Institut f\"ur Gravitationsphysik (Albert-Einstein-Institut), 30167 Hannover, Germany}
	\affiliation{Department of Physics, University of Wisconsin-Milwaukee, P.O. Box 413, Milwaukee, WI 53201, USA}
	\affiliation{Leibniz Universit\"at Hannover, 30167 Hannover, Germany}
	
	\author[0000-0001-8522-4983]{R.~P.~Breton}
	\affiliation{Jodrell Bank Centre for Astrophysics, School of Physics and Astronomy, The University of Manchester, M13 9PL, UK}
	
	\author[0000-0003-4236-9642]{V.~S.~Dhillon}
	\affiliation{Department of Physics and Astronomy, University of Sheffield, Sheffield S3 7RH, UK}
	\affiliation{Instituto de Astrof\'{i}sica de Canarias, E-38205 La Laguna, Tenerife, Spain}
	
	\author[0000-0001-5296-7035]{H.-B.~Eggenstein}
	\affiliation{Max-Planck-Institut f\"ur Gravitationsphysik (Albert-Einstein-Institut), 30167 Hannover, Germany}
	\affiliation{Leibniz Universit\"at Hannover, 30167 Hannover, Germany}
	
	\author[0000-0003-2317-1446]{J.~W.~T.~Hessels}
	\affiliation{Anton Pannekoek Institute for Astronomy, University of Amsterdam, Science Park 904, 1098 XH Amsterdam, The Netherlands}
	\affiliation{ASTRON, The Netherlands Institute for Radio Astronomy, Oude Hoogeveensedijk 4, 7991 PD Dwingeloo, The Netherlands}
	
	\author[0000-0001-6894-6044]{M.~R.~Kennedy}
	\affiliation{Jodrell Bank Centre for Astrophysics, School of Physics and Astronomy, The University of Manchester, M13 9PL, UK}
	
	\author[0000-0002-0893-4073]{M.~Kerr}
	\affiliation{Space Science Division, Naval Research Laboratory, Washington, DC 20375-5352, USA}
	
	\author[0000-0001-7221-855X]{S.~Littlefair}
	\affiliation{Department of Physics and Astronomy, University of Sheffield, Sheffield S3 7RH, UK}
	
	\author[0000-0002-2498-7589]{T.~R.~Marsh}
	\affiliation{Astronomy and Astrophysics Group, Department of Physics, University of Warwick, Coventry CV4 7AL, UK}
	
	\author[0000-0003-0245-9424]{D.~Mata~S\'anchez}
	\affiliation{Jodrell Bank Centre for Astrophysics, School of Physics and Astronomy, The University of Manchester, M13 9PL, UK}
	
	\author[0000-0002-1007-5298]{M.~A.~Papa}
	\affiliation{Max-Planck-Institut f\"ur Gravitationsphysik (Albert-Einstein-Institut), 30167 Hannover, Germany}
	\affiliation{Leibniz Universit\"at Hannover, 30167 Hannover, Germany}
	\affiliation{Department of Physics, University of Wisconsin-Milwaukee, P.O. Box 413, Milwaukee, WI 53201, USA}
	
	\author[0000-0002-5297-5278]{P.~S.~Ray}
	\affiliation{Space Science Division, Naval Research Laboratory, Washington, DC 20375-5352, USA}
	
	\author[0000-0003-1833-5493]{B.~Steltner}
	\affiliation{Max-Planck-Institut f\"ur Gravitationsphysik (Albert-Einstein-Institut), 30167 Hannover, Germany}
	\affiliation{Leibniz Universit\"at Hannover, 30167 Hannover, Germany}
	
	\author[0000-0002-4088-896X]{J.~P.~W.~Verbiest}
	\affiliation{Fakult\"at f\"ur Physik, Universit\"at Bielefeld, Postfach 100131, 33501 Bielefeld, Germany}
	\affiliation{Max-Planck-Institut f\"ur Radioastronomie, Auf dem H\"ugel 69, 53121 Bonn, Germany}

	\correspondingauthor{L.~Nieder}
	\email{lars.nieder@aei.mpg.de}
	
	\received{2019 May 24}
	\revised{2019 July 11}
	\accepted{2019 July 23}
	\published{2019 September 18}

	\begin{abstract} 
		\noindent
		The \acl{LOFAR} radio telescope discovered the $707$ Hz binary \ac{MSP} J0952$-$0607 in a targeted radio pulsation search of an unidentified \Fermi{} gamma-ray source. This source shows a weak energy flux of $F_\gamma = 2.6 \times 10^{-12}\,\text{erg}\,\text{cm}^{-2}\,\text{s}^{-1}$ in the energy range between $100\,\text{MeV}$ and $100\,\text{GeV}$. Here we report the detection of pulsed gamma-ray emission from \psr{} in a very sensitive gamma-ray pulsation search. The pulsar's rotational, binary, and astrometric properties are measured over seven years of \Fermi{}-\acl{LAT} data. For this we take into account the uncertainty on the shape of the gamma-ray pulse profile. We present an updated radio-timing solution now spanning more than two years and show results from optical modeling of the black-widow-type companion based on new multi-band photometric data taken with HiPERCAM on the Gran Telescopio Canarias on La Palma and ULTRACAM on the New Technology Telescope at ESO La Silla\footnote{Based on observations collected at the European Southern Observatory,\\Chile; programme 0101.D-0925, PI: Clark, C. J. .}. \psr{} is now the fastest-spinning pulsar for which the intrinsic spin-down rate has been reliably constrained ($\Pdot_\text{int} \lesssim 4.6 \times 10^{-21}\,\text{s}\,\text{s}^{-1}$). The inferred surface magnetic field strength of $\Bsurf \lesssim 8.2 \times 10^{7}\,\text{G}$ is among the ten lowest of all known pulsars. This discovery is another example of an extremely fast spinning black-widow pulsar hiding within an unidentified \Fermi{} gamma-ray source. In the future such systems might help to pin down the maximum spin frequency and the minimum surface magnetic field strength of \acp{MSP}.
	\end{abstract}
	

	\section{Introduction}\label{s:introduction}
	
	The \ac{LAT} on board the \Fermi{} \textit{Gamma-ray Space Telescope} \citep{atwood2009} has proven itself to be a powerful instrument in gamma-ray pulsar astronomy. Since its $2008$ launch the \ac{LAT} has been operating in an all-sky survey mode. \ac{LAT} data are used to identify promising pulsar candidates for deep, targeted radio searches and find gamma-ray pulsations in blind or follow-up searches \citep[for a review see, e.g.,][]{caraveo2014}. The $10$ year time span of the all-sky \ac{LAT} data is also useful for establishing precise pulsar-timing ephemerides of new discoveries.
	
	Radio pulsar searches targeting the sky positions of \ac{LAT} sources have been very successful in finding isolated and binary \aclp{MSP}\acused{MSP} \citep[\acsp{MSP}; e.g.,][]{ray2012}. The targeted sources are typically chosen to have three properties: (a) They are ``unassociated'', which means that the source has no plausible counterpart belonging to a known gamma-ray-emitting source class \citep[e.g.,][]{acero2015}. (b) They have curved spectra. This is parametrized in the \Fermi{}-\ac{LAT} source catalogs by the curvature significance, determined by the difference in log-likelihood between spectral models with curved spectra (e.g. a log parabola or exponentially cutoff power law) versus power-law spectra \citep{nolan2012}. For most gamma-ray pulsars, curved spectra are preferred with $>95$\% confidence \citep[e.g.,][]{abdo2013}. (c) They show only little variability in brightness over time, which is indicated in the \Fermi{} \ac{LAT} source catalogs by the variability index, the chi-squared of the monthly flux with respect to the average flux. In the \aclu{3FGL} \citep[\acs{3FGL};][]{acero2015}, only $2$ out of $136$ pulsars had variability indices corresponding to significant variability above the $99$\% confidence level. Combined, the last two properties are good indicators for gamma-ray pulsars. However, we note that the transitional \acp{MSP} \citep[for a review see, e.g.,][]{jaodand2018} are an important exception, with significant changes in gamma-ray flux associated with transitions between accretion- and rotation-powered states \citep{stappers2014,Johnson2015+J1227}.
	
	Searches following this approach continue to find pulsars by using radio observing frequencies $\nu$ above $300\,\text{MHz}$. Pulsar surveys around $350\,\text{MHz}$ are run by the Green Bank Telescope \citep[GBT;][]{stovall2014} and the Arecibo telescope \citep{cromartie2016}. The Giant Metrewave Radio Telescope searches around $607\,\text{MHz}$ \citep{bhattacharyya2013}. Another survey around $820\,\text{MHz}$ is run by the GBT \citep{ransom2011}. Finally Parkes \citep{camilo2015}, Nan\c{c}ay \citep{cognard2011} and Effelsberg \citep{barr2013} search around $1.4\,\text{GHz}$. Radio observations at higher frequencies suffer less from dispersion (dispersion delay $t_\text{d} \propto \nu^{-2}$) and scattering \citep[scattering timescale $\tau_\text{s} \propto \nu^{-4.4}$;][]{levin2016} but a pulsar's radio luminosity falls rapidly with observing frequency \citep[radio flux density $S_\nu \propto \nu^\alpha$ with spectral index $-3.0 < \alpha < -0.5$ for most known pulsars;][]{frail2016b}. At observing frequencies above $1.4\,\text{GHz}$ scattering becomes negligible away from the Galactic Center and pulsars that are bright above this frequency can be useful for Pulsar Timing Arrays \citep[e.g.,][]{verbiest2016,tiburzi2018}.
	
	However, there might be a population of steep-spectrum ($\alpha < -2.5$) radio pulsars that are most easily detectable at frequencies below $300\,\text{MHz}$. Searches by \citet{Frail2018} for steep-spectrum sources within the localization regions of unidentified \textit{Fermi}-LAT sources in continuum images from the \replaced{GMRT}{Giant Metrewave Radio Telescope} all-sky survey at $150\,\text{MHz}$ led to the discovery of six new \acp{MSP} and one normal pulsar. These detections suggest that many steep-spectrum pulsars may have been missed by high-frequency radio surveys, which favor pulsars with flatter spectra \citep{bates2013}. Additionally, some emission models suggest that pulsars' radio beams are wider at low frequencies \citep[e.g., ][]{story2007}, making pulsars whose radio beams miss our line of sight at GHz frequencies potentially detectable at lower frequencies. Low-frequency radio observations of gamma-ray pulsars can therefore provide an additional test of the viewing-angle explanation for the large number of radio-quiet pulsars discovered by the LAT \citep[e.g.][]{abdo2009d,Wu2018+EAH2}. Indeed, one emission model for the recently discovered radio-quiet \ac{MSP} PSR\,J1744$-$7619 \citep{clark2018} suggests that radio pulsations may only be detectable at low radio frequencies.
	
	\cite{pleunis2017} performed very-low-frequency pulsar searches at $115 - 155\,\text{MHz}$ with the \aclu{LOFAR} \citep[\acs{LOFAR};][]{stappers2011,haarlem2013}. This was possible due to new semi-coherent de-dispersion techniques that mitigate the smearing due to dispersion \citep{bassa2017a}. The searches targeted unassociated sources from the \ac{3FGL} catalog \citep{acero2015}. An isolated \ac{MSP}, \oLOFpsr{}, was detected first in radio and subsequently in gamma rays \added{\citep{pleunis2017}}.
	
	\cite{bassa2017b} conducted another \ac{LOFAR} survey using the same observing configuration. The $23$ targets were unassociated gamma-ray sources selected from a \Fermi{}-\ac{LAT} source list constructed from seven years of ``Pass 8'' \ac{LAT} data \citep[see][]{atwood2013}.
	
	In this survey they discovered \psr{}, a binary radio \ac{MSP} with a spin frequency of $707\,\text{Hz}$ \citep{bassa2017b}. It is in a binary system with a very-low-mass companion star ($M_\text{c} \sim 0.02 \, M_\odot$) with an orbital period of $6.42 \, \text{hr}$. \psr{} is the fastest-spinning known neutron star outside of a globular cluster: The only pulsar spinning faster ($716 \, \text{Hz}$) is \gcpsr{}, which is located in the globular cluster Terzan 5 \citep{hessels2006}. In contrast to pulsars in globular clusters, which experience significant but unknown acceleration due to the gravitational potential within the cluster \citep{prager2017}, the intrinsic spin-down rate of \psr{} can be measured directly. From this, pulsar properties like the dipole surface magnetic field strength and spin-down power can be inferred. These factors are thought to govern the poorly understood accretion and ablation processes through which binary systems containing a pulsar evolve \citep{chen2013}. Measurements of the magnetic fields of rapidly spinning pulsars are important because the origin of the low magnetic field strength of \acp{MSP} is currently unexplained, with one popular theory being that the accreted matter buries the surface magnetic field. On the other hand recent work questions if this mechanism is effective enough \citep{mukherjee2017}.
	
	To determine the pulsar properties requires precise timing solutions from frequent observations of a pulsar over several years. For some pulsar parameters (e.g. the spin frequency and spin-frequency derivative) the measurement uncertainty is directly related to the total span of observations. Furthermore, time spans shorter than one year cover less than a full cycle of the annual Roemer delay, introducing degeneracies between the spin frequency, spin-frequency derivative, and sky position. The radio-timing solution of \psr{} reported by \cite{bassa2017b} is based on observations spanning approximately $100\,\text{days}$, and thus suffers from these issues. \deleted{The detection of the optical counterpart of \psr{} allowed \cite{bassa2017b} to use the optical position in the timing solution and obtain a limit on the spin-frequency derivative of the pulsar.}

	Radio searches targeting unassociated \textit{Fermi}-LAT sources have been particularly successful at discovering ``spider pulsars'', a class of extreme binary pulsars with semi-degenerate companion stars (i.e. not neutron stars or white dwarfs). These systems are categorized as ``black widows'' if the companion star has extremely low mass ($M_\text{c} \ll 0.1 \, M_\sun$, as is the case for \psr{}) and as ``redbacks'' if the companion star is heavier ($M_\text{c} \sim 0.15 - 0.4 \, M_\sun$) \citep{roberts2013}. Optical light curves of these systems reveal that the pulsar emission heats the nearly Roche-lobe filling companion \citep{Breton2013+4MSPs}. Observations of orbitally modulated X-ray emission shows that interactions between the pulsar and companion star winds produce intra-binary shocks \citep[e.g.,][]{roberts2014}.
	
	For many spider pulsars the radio pulsations are completely absorbed by intra-binary material during parts of their orbit \citep[e.g.,][]{fruchter1988}, indicating that the companion stars are also ablated by the pulsar. At low radio frequencies these eclipses can cover a large fraction of the orbit \citep[e.g.,][]{stappers1996,archibald2009,polzin2018}, complicating radio-timing campaigns. In contrast, gamma-ray pulsations are essentially unaffected by eclipses.

	A unique value of the \ac{LAT} data is that a pulsar's discovery in gamma rays often enables the immediate measurement of the pulsar parameters over the $10$-year span in which the \ac{LAT} has been operating. \ac{LAT} data have been used to find precise timing solutions for many pulsars including radio-quiet and radio-faint pulsars \citep{ray2011,kerr2015b,clark2017}. In the case of \rbpsr{}, a strongly eclipsing redback pulsar, gamma-ray timing was essential for building a coherent timing solution, and enabled the discovery of large variations of the orbital period \citep{pletsch2015}.
	
	In this work we present the discovery and analysis of pulsed gamma-ray emission from \psr{}. The pulsar itself is very faint in gamma rays, and required novel search and timing methods with greater sensitivity. The resulting timing ephemeris extends the rotational and orbital history of \psr{} back seven years to 2011. This allows us to determine the pulsar's spin-down power and surface magnetic field strength, making it the fastest known pulsar for which these measurements can be made.

	The paper is organized as follows. In Section~\ref{s:discovery} we describe the pulsation search and detection within \ac{LAT} data. The timing analysis and resulting timing solution for \psr{} are presented in Section~\ref{s:timing}. New radio and optical observations as well as a search for continuous gravitational waves are discussed in Section~\ref{s:multiwave}. Finally, in Section~\ref{s:discussion} we discuss the implications of the results presented and we conclude in Section~\ref{s:conclusions}.

	\section{Gamma-ray Pulsation Discovery} \label{s:discovery}
	
	\subsection{Data Preparation} \label{s:dataprep}
	
	The gamma-ray source targeted by \cite{bassa2017b} resulting in the detection of the radio pulsar \psr{} and its optical counterpart (R.A. $\alpha_{\text{J2000.0}} = 09^{\rm h}52^{\rm m}08\fs319$, Decl. $\delta_{\text{J2000.0}} = -06\arcdeg07\arcmin23\farcs49$) was discovered using seven years of \ac{LAT} data, but was too faint to be included in the \ac{3FGL} catalog \citep[i.e. in four years of data;][]{acero2015}. It is included in the successive 4FGL catalog based on eight years of data as \gamsrc{} \citep{4fgl_preliminary}.
	
	To search for gamma-ray pulsations from \psr{}, we used ``Pass 8'' \citep{atwood2013} \ac{LAT} data recorded between 2008 August 4 and 2017 January 19, consisting of \texttt{SOURCE}-class photons above $500$\,MeV instead of the standard $100$\,MeV. Since the \ac{LAT}'s angular resolution for photons improves with energy ($\sim 3.6$ times higher angular resolution at $500$\,MeV compared to\,$100$\,MeV), we \added{conservatively} used $500$\,MeV to avoid \added{potential} contamination by other nearby sources not included in the \ac{3FGL} catalog\footnote{\href{https://fermi.gsfc.nasa.gov/ssc/data/analysis/documentation/Cicerone/Cicerone_LAT_IRFs/IRF_PSF.html}{https://fermi.gsfc.nasa.gov/ssc/data/analysis/documentation/Cicerone/\\Cicerone\_LAT\_IRFs/IRF\_PSF.html}}. The photons were selected using \texttt{gtselect} from the \Fermi{} Science Tools\footnote{\href{https://fermi.gsfc.nasa.gov/ssc/data/analysis/software}{https://fermi.gsfc.nasa.gov/ssc/data/analysis/software}} if they were within $10\degr$ of the celestial position of the optical counterpart to \psr{}, with a maximum zenith angle of $90\degr$. Photons were only used if the \ac{LAT} was in nominal science mode and if the rocking angle was below $52\degr$. After these cuts $114706$ \ac{LAT} photons remained for further analysis. The analysis was performed using the \texttt{P8R2\_SOURCE\_V6} \acp{IRF}.

	The sensitivity of a pulsation search can be greatly improved by weighting the contribution of each photon by its probability of having originated from the candidate pulsar \citep{bickel2008,kerr2011}. The weights are computed based on the \ac{LAT} response function and a spectral model of a point source. They are used in the search and the timing analysis for background suppression without the need for arbitrary position or \added{stronger} energy cuts.
	
	To produce the necessary spectral model we performed a binned spectral analysis with \texttt{gtlike}. We added a putative pulsar source with an exponentially cutoff power law to represent its spectrum \citep{nolan2012} fixed to the position of the pulsar's optical counterpart reported by \cite{bassa2017b}. We used the templates \texttt{gll\_iem\_v06.fits} for the Galactic diffuse emission \citep{acero2016} and \texttt{iso\_P8R2\_SOURCE\_V6\_v06.txt}\footnote{\href{https://fermi.gsfc.nasa.gov/ssc/data/access/lat/BackgroundModels.html}{https://fermi.gsfc.nasa.gov/ssc/data/access/lat/BackgroundModels.html}} for the isotropic diffuse background. The spectral analysis included all \ac{3FGL} sources within $15\degr$ of the pulsar position and the spectral parameters for point sources within $5\degr$ of the target were allowed to vary.

	For each photon within $5\degr$ of the \replaced{pulsar position}{pulsar's optical position} a probability weight $w_j$ was calculated with \texttt{gtsrcprob}. To reduce the computing cost of the search, we only included photons with $w_j > 3.1\%$. This weight cutoff value was chosen such that only $1\%$ of the expected pulsation \ac{SNR} would be lost. After applying the cutoff $N = 1354$ actual or $\sum w_j = 193.7$ ``effective'' photons remain.
	
	Upon the detection of \psr{}, we performed a dedicated spectral analysis with an extended dataset in order to enhance the pulsation significance and to model its spectral characteristics more precisely. We used the same event selection and \acp{IRF} (see above) but accepted photons without cuts on the rocking angle as this cut was found to be overly conservative\footnote{\href{https://fermi.gsfc.nasa.gov/ssc/data/analysis/documentation/Cicerone/Cicerone\_Likelihood/Exposure.html}{https://fermi.gsfc.nasa.gov/ssc/data/analysis/documentation/Cicerone/\\Cicerone\_Likelihood/Exposure.html}}. We extended the dataset to include photons between 2008 August 4 and 2018 June 21. We lowered the threshold of photon energies down to $100\,\text{MeV}$ to further constrain the spectral characteristics. We used the Preliminary LAT 8-year Point Source List\footnote{\href{https://fermi.gsfc.nasa.gov/ssc/data/access/lat/fl8y/}{https://fermi.gsfc.nasa.gov/ssc/data/access/lat/fl8y/}} (FL8Y) to construct our source model. The FL8Y source associated with the pulsar, FL8Y\,J0952.2$-$0608, was replaced by a point source fixed to the position of the detected gamma-ray pulsar. All FL8Y sources within $15\degr$ of the pulsar position were included and the spectral parameters for point sources within $5\degr$ of the pulsar were allowed to vary.
	
	We computed the residual TS map to search for non-cataloged weak gamma-ray sources in the vicinity of the pulsar. The test statistic $\text{TS} = 2 (\log \mathcal{L}(\text{source}) - \log \mathcal{L}(\text{no source}))$ quantifies how significant a source emerges from the background, where the likelihood $\mathcal{L}$ of a model with and without a source is compared \citep{nolan2012,acero2015}. Six \replaced{new}{uncatalogued} sources with $\text{TS} > 10$ \replaced{were}{($\sim3\sigma$) within $5 \degr$ of the pulsar position were found and} added to the source model. Using this new source model we reran the analysis. The result of the spectral analysis for \psr{} is shown in Table~\ref{t:spectralanalysis}. Here, we also give $\text{TS}_\text{cut}$ which is computed like TS but comparing an exponentially cutoff power-law model and a power-law model without cutoff \citep{abdo2013}.
	
	In the timing analysis we used all photons with weights $w_j > 1.5\%$, which is chosen as in the search such that $99\%$ of the \ac{SNR} remains. This leaves $N = 4642$ actual or $\sum w_j = 331.4$ effective photons.

	\begin{deluxetable}{ll}
		\tablecaption{\label{t:spectralanalysis} Spectral Parameters of \psr{}.}
		\tablecolumns{2}
		\tablehead{
			\colhead{Parameter} &
			\colhead{Value}
		}
		\startdata
		Test statistic, \text{TS} \dotfill & $147.77$ \\[0.15em]
		TS of exponential cutoff, $\text{TS}_\text{cut}$ \dotfill & $23.9$ \\[0.15em]
		Photon index, $\Gamma$ \dotfill & $0.95 \pm 0.40 \pm 0.05$ \\[0.15em]
		Cutoff energy, $E_\text{c}$ (GeV) \dotfill  & $1.62 \pm 0.55 \pm 0.01$  \\[0.15em]
		Photon flux ($10^{-9}$\,cm$^{-2}$\,s$^{-1}$)\dotfill & $2.25 \pm 0.77 \pm 0.34$ \\[0.15em]
		Energy flux $F_{\gamma}$ ($10^{-12}$\,erg\,cm$^{-2}$\,s$^{-1}$) \dotfill & $2.60 \pm 0.38 \pm 0.16$ \\[-0.05em]
		\enddata
		\tablecomments{
			Gamma-ray spectrum based on \ac{LAT} data between MJD\,$54{,}682$--$58{,}289$ over the standard energy range from $100\,\text{MeV}$ to $100\,\text{GeV}$. The first reported uncertainties are statistical, while the second uncertainties are systematic, determined by re-analyzing the data with bracketing \acp{IRF} and artificially changing the normalization of the Galactic diffuse model by $\pm6\%$, as described in \citet{abdo2013}.}
	\end{deluxetable}

	\subsection{Search} \label{s:search}
	
	For many pulsars, \ac{LAT} data covering several years of observation time are needed for significant pulsation detection \citep[e.g.][]{hou2014}. Searching for pulsations requires assigning every gamma-ray photon with the pulsar's rotational phase $\Phi$ (defined in rotations throughout the paper) at the time of emission. To do this a phase model $\Phi(t, \pvec)$ is used that depends on time $t$ and (for circular-binary pulsars) on a set of at least seven parameters $\pvec = (f, \fdot, \alpha, \delta, \Porb, x, \tasc)$. These parameters are needed to: (1) Correct the photon arrival times for the \ac{LAT}'s movement with respect to the Solar System Barycenter (sky position $\alpha$ and $\delta$). (2) In the case of a circular binary, account for the pulsar's movement around the center of mass (orbital period $\Porb$, projected semi-major axis $x$, and epoch of ascending node $\tasc$). (3) Describe the pulsar's rotation over time (spin frequency $f$ and spin-frequency derivative $\fdot$).
	
	The ephemeris obtained by timing a radio pulsar over a short interval \added{$\tobs$} often does not determine the parameters precisely enough to coherently fold the multiple years of \ac{LAT} data. For \replaced{intervals of less than a year}{$\tobs<1\,\textrm{yr}$} the spin and position parameters of the pulsar are strongly correlated (i.e., degenerate). Over longer \replaced{intervals}{$\tobs$} the uncertainties in the spin parameters scale with negative powers of \replaced{the observation time span}{$\tobs$}. The uncertainty in the orbital period scales \replaced{inversely with the observation time span}{with $\tobs^{-1}$} if \replaced{the observation time span is longer than the orbital period}{$\tobs\gg\Porb$}.
	
	Searches for binary gamma-ray pulsars are therefore computationally expensive, as a multi-dimensional parameter space must be searched with a dense grid \citep{pletsch2012}. The radio detection and timing are crucial to constrain the relevant parameter space that has to be searched to find the gamma-ray pulsations.

	Using the radio data \cite{bassa2017b} found that \psr{} is in a circular-binary orbit. Furthermore, they measured $\alpha$ and $\delta$ by identifying the companion star using optical data taken with the \ac{WFC} on the 2.5m Isaac Newton Telescope on La Palma. Barycentering the radio data according to $\alpha$ and $\delta$ obtained from the optical data resulted in an upper limit on $\fdot$ and determined $f$ more accurately. Furthermore the radio timing constrained the orbital parameters $\Porb$, $x$, and $\tasc$.

	The gamma-ray pulsation search \deleted{was performed one month after the radio discovery, and} exploited preliminary constraints from radio timing of the pulsar combined with the optical position.

	In the gamma-ray pulsation search we used the $H$ statistic \citep{dejager1989}. It combines the Fourier power from several harmonics incoherently by maximizing over the first $M$ harmonics via
	\begin{equation}
		H = \max\limits_{1 \le M \le M_\text{max}} \left( 4 - 4 M + \sum_{n=1}^{M} \mathcal{P}_n \right) \,,
	\end{equation}
	with $M_\text{max} = 20$ as suggested by \cite{dejager1989}. The Fourier power in the $n$th harmonic is given by
	\begin{equation}
	\mathcal{P}_n = \frac{1}{\kappa^2} \left| \sum_{j=1}^{N} w_j e^{-2 \pi i n\Phi(t_j)} \right|^2\,,
	\end{equation}
	with the normalization constant
	\begin{equation}
	\kappa^2 = \frac{1}{2} \sum_{j=1}^{N} w_j^2 \,.
	\end{equation}
	
	The construction of a grid for this search was done using a distance ``metric'' on the parameter space \citep{balasubramanian1996,owen1996}. This is a second-order Taylor approximation of the fractional loss in squared \ac{SNR} due to an offset from the parameters of a given signal. The metric allows one to compute analytically the density of an optimally spaced grid. This method was successfully used in the blind search (i.e., a search for a previously undetected pulsar) for the black widow \blbwpsr{} \citep{pletsch2012}.
	
	The metric components for the parameters of an isolated pulsar are given in \citet{pletsch2014}, and the additional components required to search for a binary pulsar will be described in an upcoming paper (Nieder et al. 2019, in prep.).	The grid point density computed with the metric varies throughout the parameter space. The grid density in $\alpha$ and $\delta$ increases as $f$ increases. This is also the case for the orbital parameters. In addition, for $\Porb$ and $\tasc$ the grid point density increases with the projected semi-major axis, $x$. The small $x$ typical for black-widow pulsars with their low-mass companions therefore greatly reduces the required density.
	
	In addition, when performing a harmonic-summing search, any parameter offset results in a phase offset at the $n$th harmonic that is a factor of $n$ larger than at the fundamental. To avoid this, the search grid density must be increased by a factor of $M_{\rm max}$ in each parameter. Fortunately, known gamma-ray pulsars have the most power in the first few harmonics \citep{pletsch2014}. We therefore designed the search grid to lose at most $1\%$ of the Fourier power in the fifth harmonic in each dimension. The harmonic summing was also truncated at $M_\text{max}=5$ to reduce computing cost. The required number of points in the search grid was reduced this way by a factor of $4^5$ ($\approx1000$) compared to a grid built for $M_\text{max} = 20$. This search grid was designed to be very dense since the pulsar signal was expected to be weak due to the small number of photons.
	
	Based on the distance metric we built a hypercubic grid covering the relevant parameter space in $f$, $\fdot$, $\alpha$, $\delta$, and $\Porb$. This means that the parameter space is broken down into smaller cells. The edges of these cells are parallel to the parameter axes and of equal length in each dimension as computed by the distance metric. We note that a simple hypercubic grid is sufficient because the metric is nearly diagonal (off-diagonal terms are small; Nieder et al. 2019, in prep.), and the dimensionality is low. For higher dimensional parameter spaces hypercubic grids become extremely wasteful. The projected semi-major axis and the epoch of the ascending node were known precisely enough from the radio ephemeris that no search over these parameters was necessary. In summary, we performed a grid-based search over five parameters ($f$, $\fdot$, $\alpha$, $\delta$, and $\Porb$), while keeping two parameters ($x$ and $\tasc$) fixed to the values from the radio-timing solution.
	
	The search used $2 \times 10^5$ CPU-core hours, meaning that the search would have taken $24$ years to compute on a single core. Therefore, we distributed the work in chunks over $~8000$ CPU cores of the \atlas{} computing cluster \citep{aulbert2009}, and the search took only $2$ days.

	\subsection{Detection} \label{s:detection}
	
	To ensure that the signal was inside the covered parameter space we searched over wide ranges in the highly correlated $f$ ($4\sigma$), $\alpha$, and $\delta$ ($5\sigma$ each), where $\sigma$ is the parameter uncertainty obtained from preliminary radio and optical observations. The chosen search range for $\Porb$ ($3\sigma$) was smaller because the radio-timing-derived $\Porb$ was not degenerate with the other parameters.
	
	Surprisingly, the largest $H$ statistic appeared close to the edge of our search range in $f$ and with a significant offset in $\alpha$ and $\delta$. The latter was determined to be due to an error in the initial astrometric calibration of the optical images of the optical counterpart. After the discovery of this error only the corrected $\alpha$ and $\delta$ values were published by \cite{bassa2017b}. The offset in $f$ arose from the strong correlation with $\alpha$ and $\delta$. Therefore we started another search with the same settings starting from the highest $f$ covered in the first search. The largest $H$ statistic was $H_\text{m} = 86.7$ (without refining the parameters any further) and lay well within the combined search parameter space.

	While this $H$ statistic was far larger than any other found in our search, it is not easy to estimate the statistical significance (or false-alarm probability) of the maximum value found in a dense, multi-dimensional $H$ statistic search (see Appendix~\ref{a:signif}). We therefore applied a ``bootstrapping'' procedure (described in Appendix~\ref{a:signif}) to estimate the detection significance from the search results themselves, finding a trials-corrected false-alarm probability of $P_{\rm FA} \approx 3.3\times10^{-3}$. After extending our data set to cover the extra year of data as explained in Section~\ref{s:dataprep}, and without using a weight cut (which is only introduced for computational reasons), we found that the $H$ statistic value increased to $H = 102.9$ without further refinement (i.e., in a single trial). Since no additional trials have been performed in this step, we can multiply our false-alarm probability estimate by the known single-trial false-alarm probability \citep{kerr2011} for this increase ($P_{\rm FA} = \exp(-0.3984\, \Delta H_m) = 1.6\times10^{-3}$), giving an overall false-alarm probability of $P_{\rm FA} \approx 5.3\times10^{-6}$ in the extended data set, confirming the detection.
		
	\section{Gamma-ray Timing} \label{s:timing}
	
	\subsection{Methods}

	We performed a timing analysis to measure precisely the parameters describing the pulsar's evolution over the observation time. We also allowed additional parameters to vary to test for measurable orbital eccentricity and proper motion of the binary. Instead of using a fixed search grid we use a Monte Carlo sampling algorithm to explore the parameter space around the signal parameters detected in the search. The general timing methods are also described by \cite{clark2015,clark2017}, extending the methods developed by \cite{ray2011} and \cite{kerr2015b}. We enhanced these methods with the option to marginalize over the parameters of the template pulse profile as described in detail later in this section.
	
	The starting point for the timing procedure is the construction of a template pulse profile, $\hat{g} (\Phi)$, for which we used a combination of $N_p$ symmetrical Gaussian peaks \citep{abdo2013}
	\begin{equation}
		\hat{g} (\Phi) = \left( 1 - \sum_{i=1}^{N_p} a_i \right) + \sum_{i=1}^{N_p} a_i \, g(\Phi, \mu_i, \sigma_i) \,.
	\end{equation}
	The term $a \, g(\Phi, \mu, \sigma)$ denotes a wrapped Gaussian peak with amplitude $a$, peaked at phase $\mu$ with width $\sigma$:
	\begin{equation}
		g(\Phi, \mu, \sigma) = \frac{1}{\sigma\sqrt{2\pi}} \sum_{k = -\infty}^{\infty} \exp\left(-\frac{(\Phi + k - \mu)^2}{2 \sigma^2}\right)\,.
	\end{equation}
	The phase at the first peak $\mu_1$ is chosen to be the reference phase for the template. Phases of any other peak $i$ are measured relative to the first peak as phase offset $\mu_i - \mu_1$ to avoid correlation with the overall phase. The template is fit to the weighted pulse profile obtained from the phase-folded data by maximizing over the likelihood
	\begin{equation}
		\mathcal{L}(\hat{g},\pvec) = \prod_{j=1}^{N} [w_j \hat{g}(\Phi(t_j,\pvec)) + (1 - w_j)] \,.
	\end{equation}
	The \aclu{BIC} \citep[\acs{BIC};][]{schwarz1978} is used to choose the number of peaks by minimizing
	\begin{equation}
		\text{BIC} = -2\log(\mathcal{L}(\hat{g},\pvec)) + k\log\left(\sum_{j=1}^{N} w_j\right) \,,
	\end{equation}
	where the number of free parameters in the model is denoted by $k$. Thus, adding a new parameter is penalized by $\log(\sum_{j=1}^{N} w_j)$ to avoid overfitting. The penalty factor for adding more Gaussian peaks to the template pulse profile scales with $k = 3 \times N_\text{p}$ as each peak is described by three parameters.
	
	As described by \cite{clark2017}, this template pulse profile is used to explore the multi-dimensional likelihood surface by varying the pulsar parameters with the goal to find the parameter combination that gives the maximum likelihood. We use our own implementation of the Affine Invariant Monte Carlo method described by \cite{goodman2010} to run many Monte Carlo chains in parallel for the exploration and the efficient parallelization scheme described by \cite{foreman2013}. The computations are distributed over several CPU cores. 
	
	This is repeated iteratively. Whenever a new best combination of parameters is found the template is updated using the new timing solution's phase-folded data. Usually this converges after a few iterations. Additional parameters (e.g., eccentricity) are added one after the other and the described timing procedure is restarted each time. Here again the \ac{BIC} is used to decide whether the addition of a new parameter significantly improves the pulsar ephemeris. For the timing of bright pulsars \citep[e.g.,][]{clark2017} this iterative approach is sufficient.

	For faint pulsars like \psr{}, the uncertainty in the gamma-ray pulse profile is not negligible. Using a fixed pulse profile template for weak pulsars could lead to systematic biases and underestimated uncertainties in the timing parameters. We therefore treated the template parameters in the same way as the pulsar parameters and let them vary jointly \citep[as also done in][]{an2017}.
	
	Joint variation of pulsar and template parameters results in larger but more realistic uncertainties on the pulsar parameters but should be used with a caveat. Varying pulsar parameters will always line up photons as close as possible to the same rotational phases to maximize the log-likelihood. The Monte Carlo algorithm finds combinations of parameters that lead to some photons being closer to the maximum of a peak and thus to a higher and narrower peak. But if these parameters do not describe the actual pulsar well, other photons will be shifted to phases outside the range of the peak, leading to a penalty preventing the acceptance of these parameter combinations. The joint variation of pulsar and template parameters however raises the chances of combinations that do not describe the actual pulsar well, as the peak position shifts to the phase where a combination of pulsar parameters leads to a narrow peak. This is a problem for a faint pulsar like \psr{} as the penalty factor is weaker due to the smaller amount of photons. Furthermore for a pulsar like \psr{} with two close peaks the penalty factor can be reduced by having one broader peak and one very narrow peak.
	
	To address this problem we adjusted our priors on the template parameters. As for the pulsar parameters we used uniform priors for most template parameters. For the width parameters we used log-uniform priors and constrained them to peaks broader than $5\%$ of a rotation, to disfavor extremely narrow peaks which only cover few photons, and narrower than half a rotation (full-width at half maximum $\textrm{FWHM}_i = 2\sqrt{2\log(2)} \,\sigma_i$ in the range $0.05 < \textrm{FWHM}_i < 0.5$). This led to a steadier rise in $H$  statistic over time and a pulse profile similar to what we get when folding the gamma-ray data with the updated radio-timing solution (see Section~\ref{s:radio}) reported in Table~\ref{t:timing}. In Figure~\ref{f:tim} we show $100$ pulse profile templates randomly picked from the resulting template parameter distribution.

	\subsection{Solution} \label{s:solution}
	
\begin{figure}
		\centerline{
			\hfill
			\includegraphics[width=0.99\columnwidth]{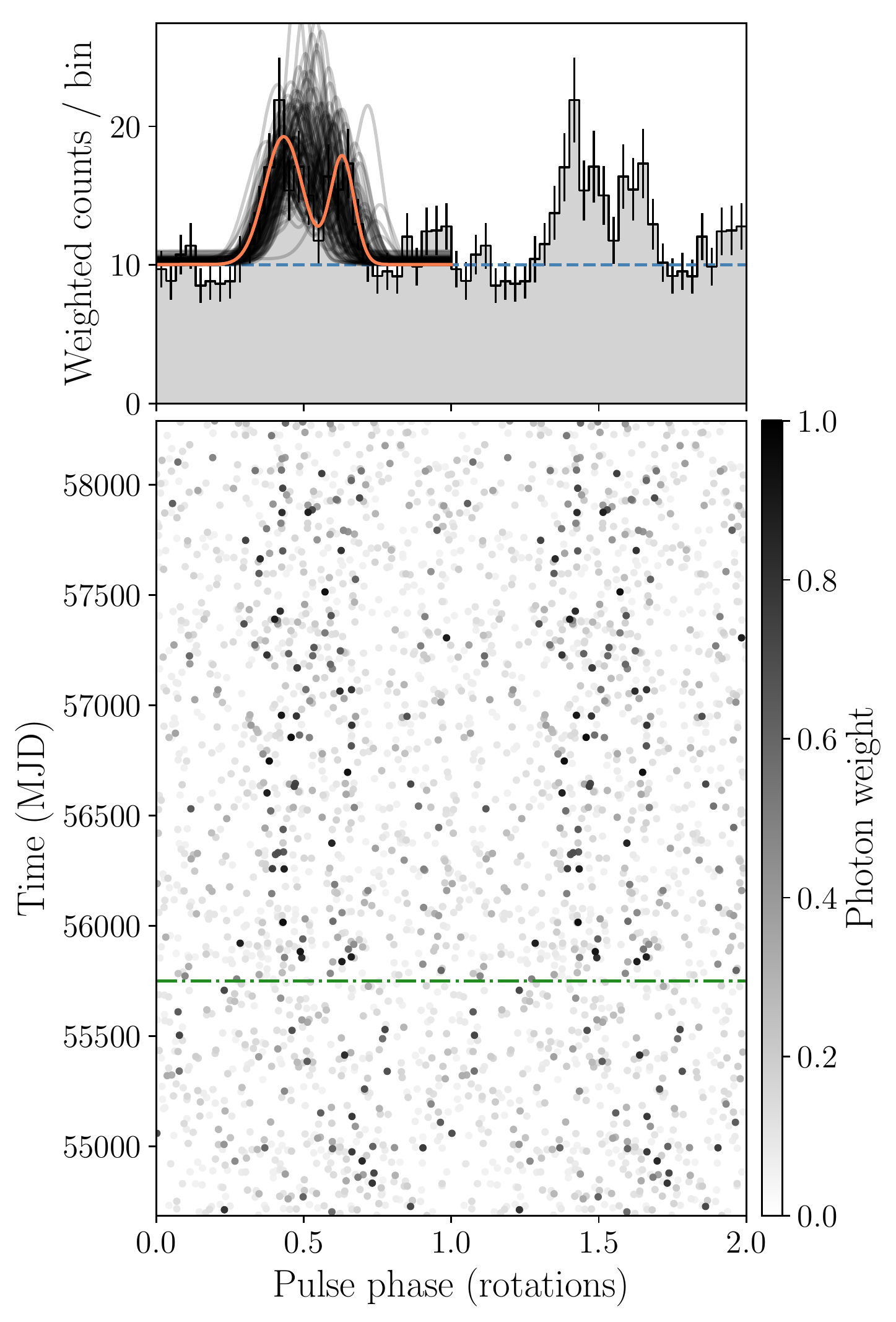}
			\hfill
		}
		\caption{\label{f:tim}
			Integrated pulse profile after MJD\,$55{,}750$ and phase-time diagram of \psr{}, showing two identical rotations for clarity.
			Top: The orange curve indicates the template with the highest \ac{BIC}. The transparent black curves illustrate $100$ representative templates randomly selected from the Monte Carlo samples after the chain stabilized. The histogram shows the weighted photon counts with $30$ bins per rotation. The dashed blue line shows the estimated background level. Bottom: Each point represents the rotational phase of a detected gamma-ray photon and its gray scale indicates the probability weight. The dashed-dotted green line denotes the start of our timing solution at MJD\,$55{,}750$.
		}
	\end{figure}

	Our timing solution is shown in Table~\ref{t:timing}. We did not find clear pulsations in the beginning of the \Fermi{} mission at MJD\,$54{,}682$ and therefore our timing solution starts at MJD\,$55{,}750$ (see Figure~\ref{f:tim}). We discuss the absence of pulsations prior to MJD\,$55{,}750$ below.
	
	The gamma-ray pulse profile is likely double peaked as the double-peaked template is favored by the \ac{BIC} over the single-peaked template. The template parameters leading to the highest likelihood are given in Table~\ref{t:timing}.
	
	All of the measured parameters are consistent with the initial published radio solution. The published values and uncertainties on $\alpha$ and $\delta$ from the optical counterpart are consistent and comparable to the ones in the gamma-ray timing solution \citep{bassa2017b}. As expected from the much longer timing baseline the uncertainties on $f$ and $\Porb$ are much smaller than in the initial radio-timing solution. Furthermore, it is possible to measure the spin-frequency derivative, $\dot{f}_\text{obs} = -2.382(8) \times 10^{-15}$ Hz s$^{-1}$. A second spin-frequency derivative, $\ddot{f}$, is clearly disfavored by the \ac{BIC}. The gamma-ray timing solution is consistent with an updated radio ephemeris based on radio data spanning $796$ days, and the parameter uncertainties are comparable or smaller (see Section~\ref{s:radio} and Table~\ref{t:timing}).

	\floattable
	\begin{deluxetable}{lll}
		\tablewidth{0.99\textwidth}
		\tablecaption{\label{t:timing} Properties of \psr{} from gamma-ray and radio timing.}
		\tablecolumns{3}
		\tablehead{
			\colhead{Parameter} &
			\colhead{Gamma-ray} &
			\colhead{Radio}
		}
		\startdata
		Span of timing data (MJD) \dotfill & $55750$\tablenotemark{a} -- $58289$ & $57759$ -- $58555$ \\[0.15em]
		Reference epoch (MJD)\dotfill & $57980$ & $57980$\\[-0.55em]
		\cutinhead{Timing Parameters}
		R.A., $\alpha$ (J2000.0)\dotfill & $09^{\rm h}52^{\rm m}08\fs322(2)$  & $09^{\rm h}52^{\rm m}08\fs32141(5)$ \\[0.15em]
		Decl., $\delta$ (J2000.0)\dotfill & $-06\arcdeg07\arcmin23\farcs51(4)$ & $-06\arcdeg07\arcmin23\farcs490(2)$ \\[0.15em]
		Spin frequency, $f$ (Hz)\dotfill & $707.3144458307(7)$ & $707.31444583103(6)$ \\[0.15em]
		Spin-frequency derivative, $\dot{f}_\text{obs}$ (Hz s$^{-1}$)\dotfill & $-2.382(8)\times 10^{-15}$  & $-2.388(4)\times 10^{-15}$ \\[0.15em]
		Dispersion measure, DM (pc cm$^{-3}$)\dotfill &  & $22.411533(11)$ \\[0.15em]
		Orbital period, $\Porb$ (day) \dotfill & $0.267461034(7)$  & $0.2674610347(5)$ \\[0.15em]
		Projected semi-major axis, $x$ (lt-s) \dotfill & $0.0626670$\tablenotemark{b}  & $0.0626670(9)$ \\[0.15em]
		Epoch of ascending node, $\tasc$ (MJD) \dotfill & $57980.4479516$\tablenotemark{b} & $57980.4479516(5)$ \\[-0.55em]
		\cutinhead{Template Pulse Profile Parameters}
		Amplitude of first peak, $\alpha_1$ \dotfill & $0.65(18)$\\[0.15em]
		Phase of first peak, $ \mu_1$ \dotfill & $0.431(39)$\\[0.15em]
		Width of first peak, $\sigma_1$ \dotfill & $0.064(23)$\\[0.15em]
		Amplitude of second peak, $\alpha_2$ \dotfill & $0.35(24)$\\[0.15em]
		Phase offset of second to first peak, $\mu_2 - \mu_1$ \dotfill & $0.198(27)$\\[0.15em]
		Width of second peak, $\sigma_2$ \dotfill & $0.040(52)$\\[-0.55em]
		\cutinhead{Derived Properties (combined results)}
		\added{Spin period, $P_\text{obs}$ (ms) \dotfill} & \multicolumn{2}{c}{1.414} \\[0.15em]
		\added{Spin-period derivative\tablenotemark{c}, $\Pdot_\text{int}$ (s\,s$^{-1}$) \dotfill} & \multicolumn{2}{c}{$4.6 \times 10^{-21}$} \\[0.15em]
		Characteristic age\tablenotemark{d}, $\tau_\text{c}$ (Gyr) \dotfill & \multicolumn{2}{c}{$4.9$} \\[0.15em]
		Spin-down power\tablenotemark{d}, $\dot{E}$ (erg s$^{-1}$) \dotfill & \multicolumn{2}{c}{$6.4 \times 10^{34}$} \\[0.15em]
		Surface $B$-field\tablenotemark{d}, $\Bsurf$ (G) \dotfill & \multicolumn{2}{c}{$8.2 \times 10^{7}$} \\[0.15em]
		Light-cylinder $B$-field\tablenotemark{d}, $B_\text{LC}$ (G) \dotfill & \multicolumn{2}{c}{$2.7 \times 10^{5}$} \\[0.15em]
		\added{Galactic longitude, $l$ (\arcdeg) \dotfill} & \multicolumn{2}{c}{$243.65$} \\[0.15em]
		\added{Galactic latitude, $b$ (\arcdeg) \dotfill} & \multicolumn{2}{c}{$+35.38$} \\[0.15em]
		\added{\citetalias{NE2001} distance, (kpc) \dotfill} & \multicolumn{2}{c}{$0.97_{-0.53}^{+1.16}$} \\[0.15em]
		\added{\citetalias{YMW16} distance, (kpc) \dotfill} & \multicolumn{2}{c}{$1.74_{-0.82}^{+1.57}$} \\[0.15em]
		\added{Optical distance, (kpc) \dotfill} & \multicolumn{2}{c}{$5.64_{-0.91}^{+0.98}$} \\[0.15em]
		\added{Gamma-ray luminosity\tablenotemark{e}, $L_{\gamma}$ (erg s$^{-1}$) \dotfill} & \multicolumn{2}{c}{$3.1 \times 10^{32} \times (d/1\,\textrm{kpc})^2$} \\[0.15em]
		\enddata
		\tablecomments{
			Numbers in parentheses are statistical $1\sigma$ uncertainties.
			The JPL DE405 solar system ephemeris has been used and times refer to TDB.
			\added{Phase $0$ is defined for a photon emitted at the pulsar system barycenter and arriving at the Solar System Barycenter at the reference epoch MJD\,$57{,}980$.}}
		\tablenotetext{\textrm{a}}{Validity range of timing solution when the data starts at MJD\,$54{,}682$.}
		\tablenotetext{\textrm{b}}{Fixed to values from radio-timing solution.}
		\tablenotetext{\textrm{c}}{Assuming no proper motion, see Section \ref{s:discussion}.}
		\tablenotetext{\textrm{d}}{Properties are derived as described in \cite{abdo2013} on the basis of the estimated intrinsic \replaced{spin-frequency derivative $\fdot_\text{int}$}{spin-period derivative $\Pdot_\text{int}$}.}
		\tablenotetext{\textrm{e}}{\added{Assuming no beaming and distance $d = 1\,\textrm{kpc}$.}}
	\end{deluxetable}
	
	It is not possible for us to confidently determine the proper motion as we find hints for and against non-zero proper motion. Allowing proper motion to vary jointly with the template parameters results in a significantly improved $H$ statistic, log-likelihood, and \ac{BIC}. The timing analysis sets the $95\%$ confidence region on proper motion to $\mu_{\alpha} \cos\delta \in [-27.4, -1.9] \,\textrm{mas}\,\textrm{yr}^{-1}$ and $\mu_{\delta} \in [-23.0, 19.1]\,\textrm{mas}\,\textrm{yr}^{-1}$. The most likely total proper motion $\mu_{\textrm{t}} = \sqrt{ \mu_{\alpha}^2 \cos^2\delta + \mu_{\delta}^2 }$ is $14.8\,\textrm{mas}\,\textrm{yr}^{-1}$ with a $95\%$ upper limit of $25.3\,\textrm{mas}\,\textrm{yr}^{-1}$. Typically, however, it is assumed that the $H$ statistic rises linearly with exposure time and non-zero proper motion resulting from this timing analysis leads to a bumpier rise in the $H$ statistic over time. This indicates that the proper motion resulting from our analysis might not be correct. Keeping the template fixed to the template parameters found by folding the gamma-ray data with the radio ephemeris results in a $95\%$ confidence region on proper motion consistent with zero. Zero proper motion is also favored by the \ac{BIC}. The same is found when using a single-peaked profile in the timing analysis and varying the template parameters jointly.
	
	The upper limit on proper motion corresponds to a transverse velocity of $v_\text{t} = \mu_\text{t} \, d = 120 \, \text{km\,s}^{-1} \times (d/1\,\text{kpc})$. This results in high, but not unrealistic transverse velocities when using the distances inferred from the dispersion measure ($d = 0.97\,\text{kpc}$ \citep[][hereafter \citetalias{NE2001}]{NE2001} or $d = 1.74\,\text{kpc}$ \citep[][hereafter \citetalias{YMW16}]{YMW16}). As $90\%$ of the known \acp{MSP} in the ATNF Pulsar Catalogue\footnote{\label{n:atnf}\href{http://www.atnf.csiro.au/research/pulsar/psrcat}{http://www.atnf.csiro.au/research/pulsar/psrcat}} \citep{manchester2005} show transverse velocities below $200\,\text{km\,s}^{-1}$ the proper motion upper limit is unrealistic for the higher distances predicted by the optical observations ($4.7$--$6.6\,\text{kpc}$; see Section~\ref{s:discussion}).
	
	Unsurprisingly, we were unable to detect a significant timing parallax. The maximum parallax time delay for the above-mentioned distance estimates is $\Delta t_{\pi,\max} \approx (500 \, \text{lt-s})^2/(2d) \sim 1\,\mu \text{s}$. In comparison the resolution with which we can measure the arrival time of the pulse is $\Delta \mu_1 / f \approx 61 \, \mu\text{s}$.
	
	A circular orbit is clearly favored over an eccentric orbit by the \ac{BIC}. The $95\%$ upper limit on eccentricity is set to $e<0.004$.
	
	The missing pulsations before MJD\,$55{,}750$ seem odd as the tracks are clearly visible later in the mission (Figure~\ref{f:tim}). As the pulsar is not very bright one explanation might be Poisson variations in the flux leading to the loss of pulsations for a few hundred days. Possible pulsations before this period might be too weak to be picked up again as the phase uncertainty grows quickly outside the timing span. At the start of the mission (MJD\,$54{,}682$) the phase uncertainty is $\sim 0.6$ rotations, which could be a plausible explanation for loss of coherence.
	
	In order to understand the nature of the non-detection of gamma-ray pulsations before MJD\,55750, we \replaced{searched for uncatalogued sources around \psr{}. We found $6$ uncatalogued sources with TS $> 10$ ($\sim3$ sigma) within $5 \degr$ of \psr{}. These sources were then included in the source model for subsequent analysis (see Section~\ref{s:dataprep}). To measure}{measured} the gamma-ray flux of \psr{} over time \replaced{we slid}{by sliding} a $750$-day-long window in steps of $50$ days over the \ac{LAT} data. In each of these steps we calculated the gamma-ray flux of \psr{} over the $750$ days width of the window, which allowed us to measure the spectral parameters with reasonable precision. We found that the flux of \psr{} is lower in the beginning of the \Fermi{} mission but the lower fluxes agree with the flux uncertainties from the full time span. The TS values follow the same trend as the gamma-ray fluxes in the sliding windows.
	
	The gamma-ray source is too faint to test it unambiguously for variability. The windows need to cover $750$ days to keep statistical precision. But that leaves only five independent time bins to calculate the variability index with Equation~($4$) from \cite{abdo2010b}. The variability index computed with these five bins is $7.18$ with $4$ degrees of freedom, which is below the $99\%$ confidence level of $13.277$.
	
	We also checked if the smaller $35\arcdeg$ rocking angle used during the first year of the \Fermi{} mission decreases the pulsation significance. However, the small rocking angle is actually favorable as the exposure for \psr{} is \added{$\sim 20\%$} higher in the beginning of the mission.
	
	Variations of the orbital period might be another reasonable explanation for the loss of clear pulsations. Such orbital-period variations have been measured for several spider pulsars, e.g. for the original black-widow pulsar \fbwpsr{} \citep{arzoumanian1994}. Nevertheless the penalty for adding orbital-period derivatives led to an increase in the \ac{BIC}. Similarly, no significant semi-major-axis derivative was found.

	\section{Multi wavelength}\label{s:multiwave}
	
	\subsection{Updated Radio Timing}\label{s:radio}
	
	\deleted{\citet{bassa2017b} presented a timing solution for \psr{} based on observations with the \ac{LOFAR} \acp{HBA} over a time baseline of approximately $100$ days. Due to this short time baseline, the astrometric and rotational parameters were degenerate, and a constraint on $\fdot$ was only possible by fixing the pulsar position to that of the optical counterpart. }Observations of \psr{} with \ac{LOFAR} have been ongoing using an identical observational setup as in \citet{bassa2017b}, namely a single tied-array beam formed from the \acp{HBA} of the central $23$ LOFAR \citep{haarlem2013} core stations, using $78\,\text{MHz}$ of bandwidth at a central frequency of $149\,\text{MHz}$. Before $2018$ May, several $5\,\text{min}$ integrations were obtained at each observing epoch; after that the integration times were increased to $20\,\text{min}$. These observations were obtained at a roughly monthly cadence. As described in \citet{bassa2017b}, these observations were coherently de-dispersed, folded with \textsc{dspsr} \citep{straten2011}, and analyzed using tools in the \textsc{psrchive} software suite \citep{hotan2004} and the \textsc{tempo2} pulsar-timing software \citep{edwards2006,hobbs2006}.
	
	The phase-connected timing solution from \citet{bassa2017b} was improved by using all \ac{LOFAR} \ac{HBA} observations that used $78\,\text{MHz}$ of bandwidth (hence excluding the discovery and initial follow-up observations which used half the bandwidth). Pulse time-of-arrival (TOA) measurements were obtained by referencing pulse profiles of eight frequency channels per observation to a single analytic pulse profile template. This procedure presumes that our data are not sensitive to pulse profile shape variations with frequency, which was double-checked through inspection of the difference profiles of the top and bottom parts of the bandpass: no significant structures were detected. The analytic pulse profile was created using the \textsc{psrchive} \citep{straten2012} package \textsc{paas} and was constructed from five von Mises functions that were fitted to the integrated body of observations and fully modeled any detectable pulse shapes. The resulting timing solution extends the timing baseline to $2.2$ years and breaks the degeneracy between the astrometric and rotational parameters (see Table~\ref{t:timing}). Upon inspection of the data, a new covariance was detected, namely, between a significant ($>4\sigma$) decrease in the dispersion measure of this pulsar (which was found to be decreasing by $5\times 10^{-5}\,\text{pc}\,\text{cm}^{-3}\,\text{yr}^{-1}$) and the spin period. Notwithstanding the significance of this decrease, the strong anticorrelation of this parameter with the pulse period suggests an underestimate of its measurement significance, which is commonly found in pulsar-timing analyses \citep[e.g.,][]{coles2011}, particularly in non-periodic parameters such as linear gradients in dispersion measure. Consequently this decrease was not included in our present analysis, but future monitoring to allow more robust disentanglement of the spin period and the dispersion measure variability is warranted. We find no evidence for radio eclipses in the six \ac{LOFAR} observations with orbital phases between $0.15<\phi_\mathrm{orb}<0.35$. Using the TOAs from this orbital phase range we set a $3\sigma$ upper limit on time delays due to additional dispersion of $\Delta t < 2.3$\,$\mu$s, and hence $\Delta \mathrm{DM}<1.2\times10^{-5}$\,pc\,cm$^{-3}$.

	\subsection{Optical Photometry}\label{s:optical}
	
	\citet{bassa2017b} presented an $r^\prime$-band light curve of the optical companion to \psr{} taken by the \ac{WFC} on the 2.5m Isaac Newton Telescope on La Palma. The orbital light curve features a single maximum peaking at $r^\prime \approx 22$ at the pulsar's inferior conjunction, interpreted as being due to the pulsar heating the inside face (the ``dayside'') of a tidally locked companion. \cite{bassa2017b} modeled this light curve with the \texttt{Icarus} package \citep{Breton2012+Icarus}, finding that \psr{} is likely to have an inclination angle $i\sim40\degr$, but the lack of color information precluded a robust estimate of other system parameters (e.g. companion temperature, heating, companion radius).
  
  	\floattable
		\begin{deluxetable}{lcccccc}
  			\tablewidth{0.99\textwidth}
  			\tablecaption{\label{t:opticaldata} New optical photometry of the companion of \psr{}.}
  			\tablecolumns{4}
  			\tablehead{
  				\colhead{Night beginning} &
  				\colhead{Instrument+Telescope} &
  				\colhead{Filters} &
  				\colhead{$\phiorb$} &
  				\colhead{Airmass} &
  				\colhead{Seeing} &
          \colhead{Photometric}
  			}
  			\startdata
  			2018 June 03 & ULTRACAM+NTT & $u_s$, $g_s$, $i_s$ & $0.64$--$1.09$ & $1.1$--$2.1$ & $1.0\arcsec$--$2.0\arcsec$ & yes\\
  			2018 June 04 & ULTRACAM+NTT & $u_s$, $g_s$, $i_s$ & $0.37$--$0.71$ & $1.1$--$1.6$ & $1.0\arcsec$--$3.0\arcsec$ & no\\
  			2019 January 12 & HiPERCAM+GTC & $u_s$, $g_s$, $r_s$, $i_s$, $z_s$ & $0.77$--$0.92$ & $1.25$--$2.0$ & $<1.5\arcsec$ & yes\\
  			2019 January 13\tablenotemark{a} & HiPERCAM+GTC & $u_s$, $g_s$, $r_s$, $i_s$, $z_s$ & $0.37$--$0.72$ & $1.25$--$2.0$ & $1.5\arcsec$--$2.0\arcsec$ & no\\
  			2019 March 02\tablenotemark{b} & ULTRACAM+NTT & $u_s$, $g_s$, $i_s$ & $0.91$--$1.29$ & $1.1$--$1.6$ & $0.8\arcsec$--$1.2\arcsec$ & no\\
  			2019 March 03 & ULTRACAM+NTT & $u_s$, $g_s$, $i_s$ & $0.72$--$0.88$ & $1.2$--$1.4$ & $1.2\arcsec$--$2.4\arcsec$ & no \\
  			 &  &  & $1.16$--$1.72$ & $1.1$--$1.9$\\
  			\enddata
  			\tablecomments{
  				Orbital phases are in fractions of an orbit, with $\phiorb = 0$ corresponding to the pulsar's ascending node. The ULTRACAM data from $2018$ were taken as a series of $20$s exposures in $g_s$ and $i_s$, and $60$s in $u_s$. The 2019 ULTRACAM observations were taken with $10$s exposures in $g_s$ and $i_s$, and $30$s in $u_s$. The HiPERCAM data cover $u_s, g_s, r_s, i_s$, and $z_s$ simultaneously with exposure times of 60s in $u_s, g_s, r_s$, and 30s in $i_s$ and $z_s$.
  			}
  			\tablenotetext{\textrm{a}}{During an episode around $\phiorb = 0.6$ seeing reached over $2.3\arcsec$ and $20$ exposures had to be removed.}
  			\tablenotetext{\textrm{b}}{We removed several frames due to intermittent clouds during the observations when the transmission dropped to nearly zero.}
		\end{deluxetable}

	To more fully investigate the optical counterpart to \psr{}, we obtained multi-color photometry using ULTRACAM \citep{Dhillon2007+ULTRACAM} on the 3.58m New Technology Telescope (NTT) at ESO La Silla, and HiPERCAM \citep{Dhillon2016+HiPERCAM,dhillon2018} on the 10.4m Gran Telescopio Canarias (GTC) on La Palma. The observation specifics are given in Table~\ref{t:opticaldata}.

	These data were calibrated and reduced using the ULTRACAM\footnote{\href{http://deneb.astro.warwick.ac.uk/phsaap/software/ultracam/html/}{http://deneb.astro.warwick.ac.uk/phsaap/software/ultracam/html/}} and HiPERCAM\footnote{\href{http://deneb.astro.warwick.ac.uk/phsaap/hipercam/docs/html/}{http://deneb.astro.warwick.ac.uk/phsaap/hipercam/docs/html/}} software pipelines. Standard CCD calibration procedures were applied using bias and flat field frames taken during each run.

	We extracted instrumental magnitudes using aperture photometry, and performed ``ensemble photometry'' \citep{Honeycutt1992+EnsemblePhot} to correct for airmass effects and varying transparency. Magnitudes in $g_s, r_s, i_s$, and $z_s$ \footnote{ULTRACAM and HiPERCAM use identical higher-throughput versions of the SDSS filter set, which we refer to as {\em Super-SDSS filters}: $u_s$, $g_s$, $r_s$, $i_s$, and $z_s$ \citep{dhillon2018}.} were calibrated using comparison stars chosen from the Pan-STARRS1 \citep{PanSTARRS} catalog, after fitting for a color term accounting for differences between our filter sets and the Pan-STARRS1 filters. The HiPERCAM $u_s$ observations were flux calibrated using zero-points derived from observations of two Sloan Digital Sky Survey (SDSS) standard stars \citep{Smith2002+SDSS} taken on 2019 January 11. The resulting HiPERCAM magnitudes for three nearby stars to \psr{} were used to flux calibrate the ULTRACAM $u_s$ data. Finally, the airmass- and ensemble-corrected count rates ($C$) were converted to AB flux densities according to our measured zero-point counts in each frame ($C_0$) by $S_{\rm AB} = 3631\, (C/C_0)$\,Jy.

	\subsection{Optical Light-Curve Modeling}
	
	As in \citet{bassa2017b}, the \texttt{Icarus} software was used to estimate parameters of the binary system. To do this, we performed a Bayesian parameter estimation using the nested sampling algorithm \texttt{MultiNest} \citep{MultiNest} via the Python package \texttt{PyMultiNest} \citep{PyMultiNest}. \texttt{Icarus} produces model light curves by computing a grid of surface elements covering the companion star, and calculating and summing the projected line-of-sight flux from each element. Here the flux from each surface element was computed by integrating spectra from the G\"ottingen Spectral Library models of \citet{Husser2013+Atmo}.

	In these fits we assumed that the companion star is tidally locked to the pulsar, and varied the following parameters: the companion star's ``nightside'' temperature $(T_\text{n})$; the ``irradiating temperature'' ($T_\text{irr}$ defined such that the dayside temperature $T_\text{d}^4 = T_\text{irr}^4 + T_\text{n}^4$, under the assumption that the pulsar's irradiating flux is immediately thermalized and re-radiated, and therefore simply adds to the companion star's intrinsic flux at each point on the surface, as in \citealt{Breton2013+4MSPs}); the binary inclination angle $(i)$; the Roche-lobe filling factor ($f_{\rm RL}$, defined as the ratio between the companion's radius towards the pulsar and the inner Lagrange point (L1) radius); the distance modulus $(\mu = 5 \log_{10}(d) - 5)$, with distance $d$ in pc; and the mass of the pulsar $(M_{\rm psr})$. At each point, the companion mass $(M_{\rm c})$ and mass ratio $(q = M_{\rm psr}/M_{\rm c})$ were derived from the binary mass function according to the timing measurements of $\Porb$ and $x$ presented in Table~\ref{t:timing}. We also marginalize over interstellar extinction and reddening, parameterized by the $E(B-V)$ of \citet{Green2018+Av}, scaled using the coefficients given therein for Pan-STARRS1 filter bands. We adopted a Gaussian prior for $E(B-V)$ (truncated at zero), using the value from \citet{Green2018+Av} for $d>1\,\text{kpc}$ in the direction of \psr{}, $E(B-V) = 0.065 \pm 0.02$, found by fitting the line-of-sight dust distribution using the apparent magnitudes of nearby main-sequence stars in the Pan-STARRS1 catalog. We adopted uniform priors on the remaining parameters (and uniform in $\cos i$), with $M_{\rm psr}$ and $f_{\rm RL}$ limited to lie within $1.2 < M_\text{psr} < 2.5\,M_{\odot}$, and $0.1 < f_\text{RL} < 1$. Temperatures $T_{\rm n}$ and $T_{\rm d}$ were constrained to lie within the range covered by the atmosphere models, $2300 < T < 12000\,\text{K}$.

	At each point in the sampling, \texttt{Icarus} computed model light curves in each band. To account for remaining systematic uncertainties in the flux calibration, extinction, and atmosphere models, the model light curve in each band was re-scaled at each parameter location to maximize the penalized chi-squared log-likelihood. Overall calibration offsets were allowed for each band, and penalized by a zero-mean Gaussian prior on the scaling factor in each band with a width of $0.1$\,mag (a conservative estimate based on our calibration to the Pan-STARRS1 magnitudes). We also allowed small offsets between the calibrations for each ULTRACAM run and the HiPERCAM observations, which we penalized with an additional Gaussian prior with width $0.05$\,mag (also a conservative estimate from the differences in magnitudes of comparison stars in the field of view on each night). In initial fits, our best-fitting model resulted in a reduced chi-squared greater than unity. We therefore also re-scaled the uncertainties in each band to maximize the (re-normalized) log-likelihood at each point in the sampling. We also found that the fit improved substantially when we fit for a small orbital phase offset. Such orbital phase offsets are often seen in the optical light curves of black-widow pulsars and have been interpreted as being due to asymmetric heating from the pulsar, which could be caused by reprocessing of the pulsar wind by an intra-binary shock \citep[e.g.,][]{Sanchez2017+Bduct}.

\begin{figure*}
		\centerline{
			\hfill
			\includegraphics[width=0.99\textwidth]{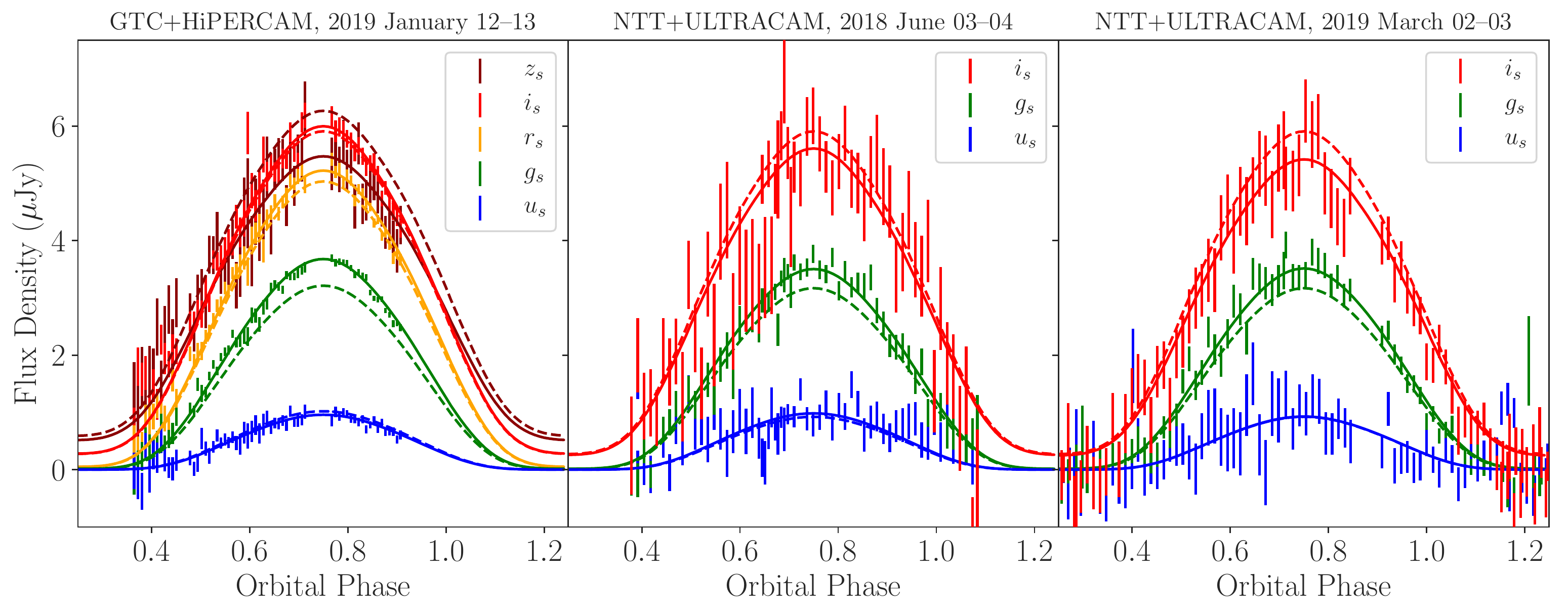}
			\hfill
		}
		\caption{\label{f:Icarus_LC} Optical light curve of the companion to \psr{}, phased using the gamma-ray timing ephemeris. For clarity, the HiPERCAM and ULTRACAM fluxes have been combined into $180$ and $300$\,s time bins, respectively, via weighted average. The unbinned data were used for the light-curve model fitting. Dashed and solid curves show the flux in each band as predicted by the best-fitting \texttt{Icarus} model before and after allowing for uncertainties in the flux calibrations (see text), respectively.}
	\end{figure*}

\begin{figure*}
		\centerline{
			\hfill
			\includegraphics[width=0.99\textwidth]{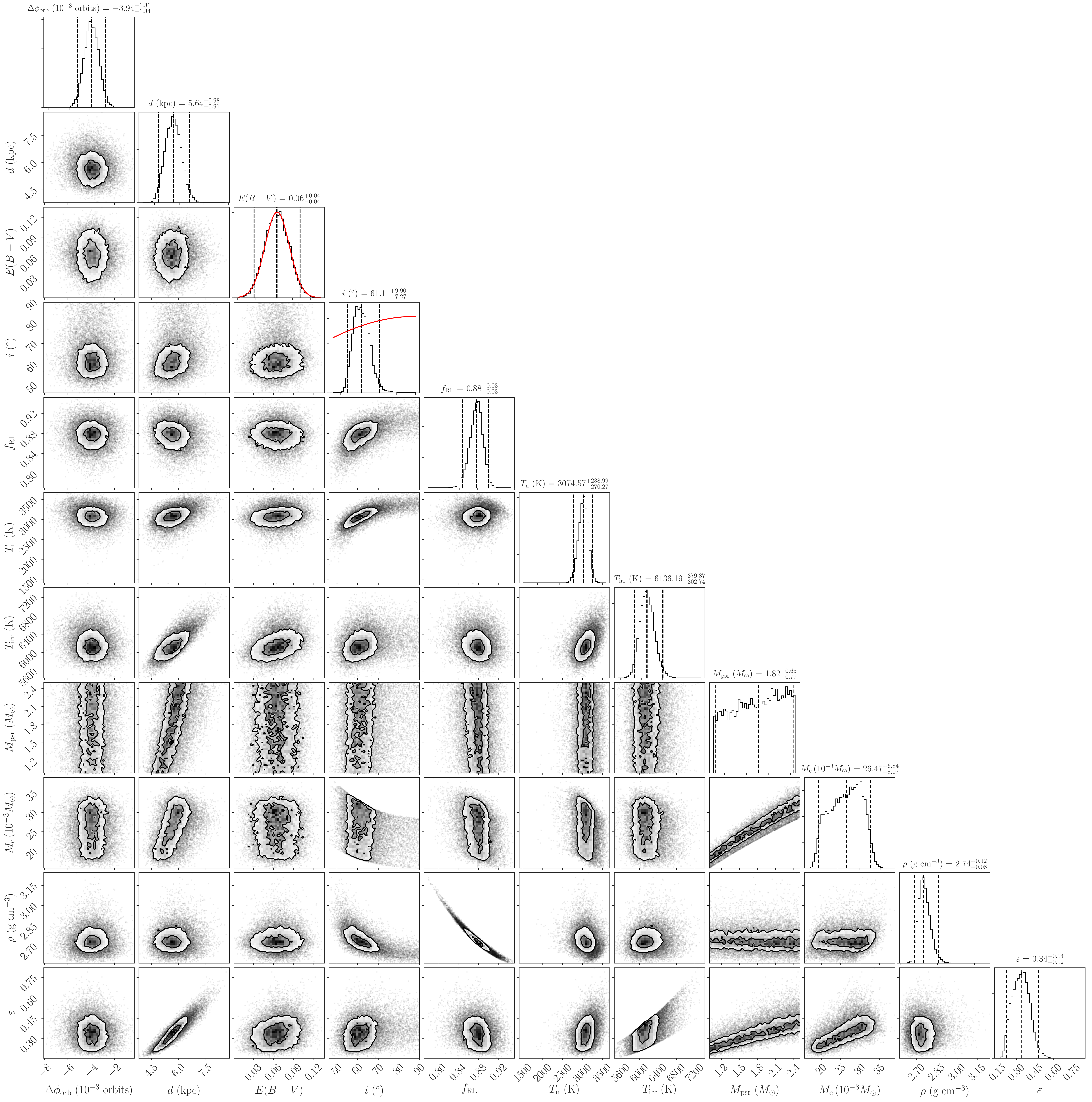}
			\hfill
		}
		\caption{\label{f:Icarus_MCMC} Posterior distributions for optical light-curve modeling parameters. The last three parameters (companion mass $M_{\rm c}$, volume-averaged density $\rho$ and heating efficiency $\varepsilon$) were derived from the values of the other fit parameters and the gamma-ray timing ephemeris. Dashed vertical lines on histograms indicate the posterior mean and $95$\% confidence interval. Where non-uniform priors were assumed, these are shown by red curves on the one-dimensional histograms. Contour lines indicate $1\sigma$ and $2\sigma$ confidence regions, with individual samples outside these areas shown as points weighted by their posterior probability.}
	\end{figure*}

	The best-fitting light-curve model is shown in Figure~\ref{f:Icarus_LC}, with posterior distributions for the fit parameters shown in Figure~\ref{f:Icarus_MCMC}.

	\subsection{Search for Continuous Gravitational Waves}
	
	We carried out a search for near-monochromatic continuous gravitational waves phase locked at twice the pulsar rotation phase for the source \psr{} using data from the first and second runs (O1\footnote{\href{https://doi.org/10.7935/K57P8W9D}{https://doi.org/10.7935/K57P8W9D}} and O2\footnote{\href{https://doi.org/10.7935/CA75-FM95}{https://doi.org/10.7935/CA75-FM95}}) of the two Advanced LIGO detectors \citep{vallisneri2015LOSC}. The observation period spans {\DataDaysTot} from 2015 September to 2017 August and comprises {\DataDaysH} ({\DataDaysL}) of data from the Hanford (Livingston) detector.
	
	We employ the coherent multi-detector detection statistic $\twoF$ \citep{jaranowski1998, cutler2005} that we implemented in the \textsc{LIGO-Lalsuite} library\footnote{\href{https://git.ligo.org/lscsoft/lalsuite/}{https://git.ligo.org/lscsoft/lalsuite/}}. $\twoF$ is the log-likelihood maximized over the amplitude parameters $h_0, \cos\iota, \psi$ and $\Phi_0$ for a near-monochromatic\footnote{The signal is not strictly monochromatic because of the measured non-zero spin-frequency derivative.} gravitational wave signal with given frequency and frequency-derivative values, from a source in a binary at a given sky position and with given orbital parameters, in Gaussian noise. $h_0$ is the intrinsic gravitational wave amplitude at the detector, $\iota$ the angle between the total angular momentum of the pulsar and the line of sight to it from Earth, $\psi$ is the gravitational wave polarization angle and $\Phi_0$ the signal phase at a nominal reference time. In this search we assume the gravitational wave frequency and frequency derivatives equal to twice the values measured for the pulsar rotation frequency and its derivatives. In Gaussian noise the detection statistic $\twoF$ follows a $\chi^2$-distribution with 4 degrees of freedom and non-centrality parameter equal to 0: the expected value is $\mu = 4.0$, and the standard deviation is $\sigma = 2\sqrt{2}$. If a signal is present, the non-centrality parameter is proportional to the square of the intrinsic gravitational wave amplitude at the detector, $h_0$, and to the total observation time.
	
	The search yields the value $\twoF = \twoFcand$, which is well within the bulk of the distribution consistent with a null result. Based on the measured value of the detection statistic, we set a frequentist $95\%$ upper limit on the intrinsic gravitational wave amplitude, $\hUL$, following a now standard procedure first developed by some of us \citep{abbott2004}. $\hUL$ is the smallest intrinsic gravitational wave amplitude such that 95\% of the population of signals that could be emitted by \psr{}\footnote{The possible signals span uniformly distributed values of $-1 \leq \cos\iota \leq 1$ and of ${0} \leq \psi\leq {2\pi}$.} would yield a detection statistic value greater than the measured one, $\twoF = \twoFcand$. We find  $\hUL = \hULval$. The uncertainty on this upper limit is $\sim \calberr$, including instrument calibration errors \citep{cahillane2017}.

	\section{Discussion}\label{s:discussion}
	
	The pulsar's spin period is defined as $P = 1/f$ and the spin-period derivative is $\dot{P} = -\fdot/f^2$. The observed spin period for \psr{} from gamma-ray and radio timing is $P_\text{obs} = 1.414\,\text{ms}$ and the observed spin-period derivative is $\Pdot_\text{obs} = 4.76 \times 10^{-21}\,\text{s}\,\text{s}^{-1}$.
	
	The intrinsic spin-period derivative $\Pdot_\text{int}$ can be estimated from the observed value $\Pdot_\text{obs} = \Pdot_\text{int} + \Pdot_\text{Gal} + \Pdot_\text{Shk}$. $\Pdot_\text{Gal}$ represents the part of the spin-period derivative caused by the relative Galactic acceleration \citep[differential Galactic rotation and acceleration due to the Galactic gravitational potential; e.g.,][]{damour1991,nice1995}, while $\Pdot_\text{Shk}$ accounts for the Shklovskii effect due to non-zero proper motion \citep{shklovskii1970}. \deleted{To estimate these effects we use the dispersion measure distances $d = (0.97,1.74)\,\text{kpc}$ \citep{bassa2017b} predicted by the (\citetalias{NE2001}, \citetalias{YMW16}) models. Here and in the following the brackets indicate the two assumed distances.} \added{Both contributions, $\Pdot_\text{Gal}$ and $\Pdot_\text{Shk}$, depend on the distance $d$ to the pulsar.}
	
\begin{figure}
		\centerline{
			\hfill
			\includegraphics[width=0.99\columnwidth]{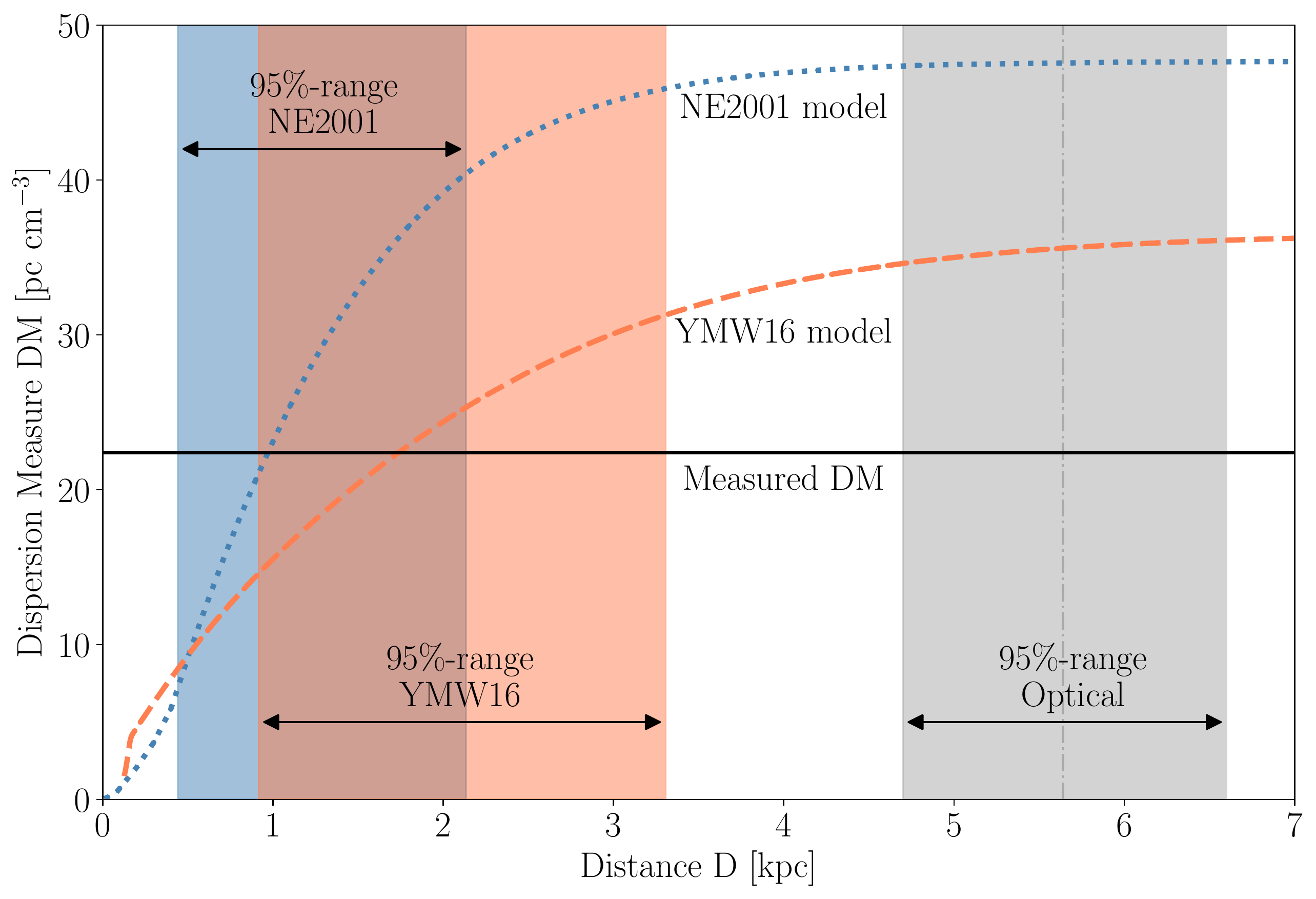}
			\hfill
		}
		\caption{\label{f:dmdistance}
			\added{
				Dispersion measure versus distance from the \citetalias{NE2001} and \citetalias{YMW16} models at the sky position of \psr{}. For the measured $\textrm{DM} = 22.4 \,\textrm{pc}\,\textrm{cm}^{-3}$ (black, horizontal line) the \citetalias{NE2001} model (dotted, blue line) and the \citetalias{YMW16} model (dashed, orange line) predict distances of $0.97\,\textrm{kpc}$ and $1.74\,\textrm{kpc}$, respectively. The $95\%$ confidence regions around those values are calculated as $120\%$ (\citetalias{NE2001}) and $90\%$ (\citetalias{YMW16}) ``relative'' errors on the predicted values \citep{YMW16}. To illustrate the discrepancy with these distance predictions, the $95\%$ confidence region from the optical modeling is shown. The vertical, dashed-dotted line indicates the distance favored by the optical modeling.
			}
		}
	\end{figure}

	\added{The distance to \psr{} is uncertain. The measured DM can be used to estimate the distance using Galactic electron-density models. The \citetalias{NE2001} model predicts $0.97_{-0.53}^{+1.16}\,\textrm{kpc}$, while the \citetalias{YMW16} model predicts $1.74_{-0.82}^{+1.57}\,\textrm{kpc}$. The uncertainties represent the $95\%$ confidence regions \citep{YMW16}. The model predictions of the DM as a function of $d$ in the direction of the pulsar's sky position are shown in Figure~\ref{f:dmdistance}. The models saturate at DM values that differ by $\sim30\%$ indicating the challenge and difficulty modeling the Galactic electron density. Still the distance predictions are consistent within the large uncertainty. On the other hand, the distance derived from optical modeling is $5.64_{-0.91}^{+0.98}\,\textrm{kpc}$. This disagrees strongly with both DM distances and suggests that both DM models are overestimating the electron density in the direction of \psr{}. The distance discrepancy is discussed in more detail below.}
	
	The estimated Galactic contribution is $\Pdot_\text{Gal} = (1.7,2.2,3.6) \times 10^{-22}\,\text{s} \, \text{s}^{-1}$\added{ for the distance estimates $d = (0.97,1.74,5.64)\,\text{kpc}$}. For the Shklovskii effect we then find the $95\%$ confidence region to $\Pdot_\text{Shk} \in ([0, 2.1],[0, 3.8]) \times 10^{-21}\,\text{s}\, \text{s}^{-1}$ from the proper motion $95\%$ confidence region  (see Section~\ref{s:solution}) \added{ and for the (\citetalias{NE2001}, \citetalias{YMW16}) distances. The resulting $95\%$ confidence region on $\Pdot_\text{Shk}$ for the optical distance exceeds past $\Pdot_\text{obs}$.} Thus we \added{only} constrain the intrinsic spin-frequency derivative (at $95\%$ confidence) to $\Pdot_\text{int} \in [2.44,4.59] \times 10^{-21} \, \text{s} \, \text{s}^{-1}$ for the \citetalias{NE2001} model and $\Pdot_\text{int} \in [0.69,4.54] \times 10^{-21} \, \text{s} \, \text{s}^{-1}$ for the \citetalias{YMW16} model. In the following, we conservatively \replaced{assumed}{assume} zero proper motion (i.e. $\Pdot_\text{Shk} = 0$) and used the fastest possible spin-down rate, $\Pdot_\text{int} = 4.6 \times 10^{-21} \, \text{s} \, \text{s}^{-1}$.
	
\begin{figure}
		\centerline{
			\hfill
			\includegraphics[width=0.99\columnwidth]{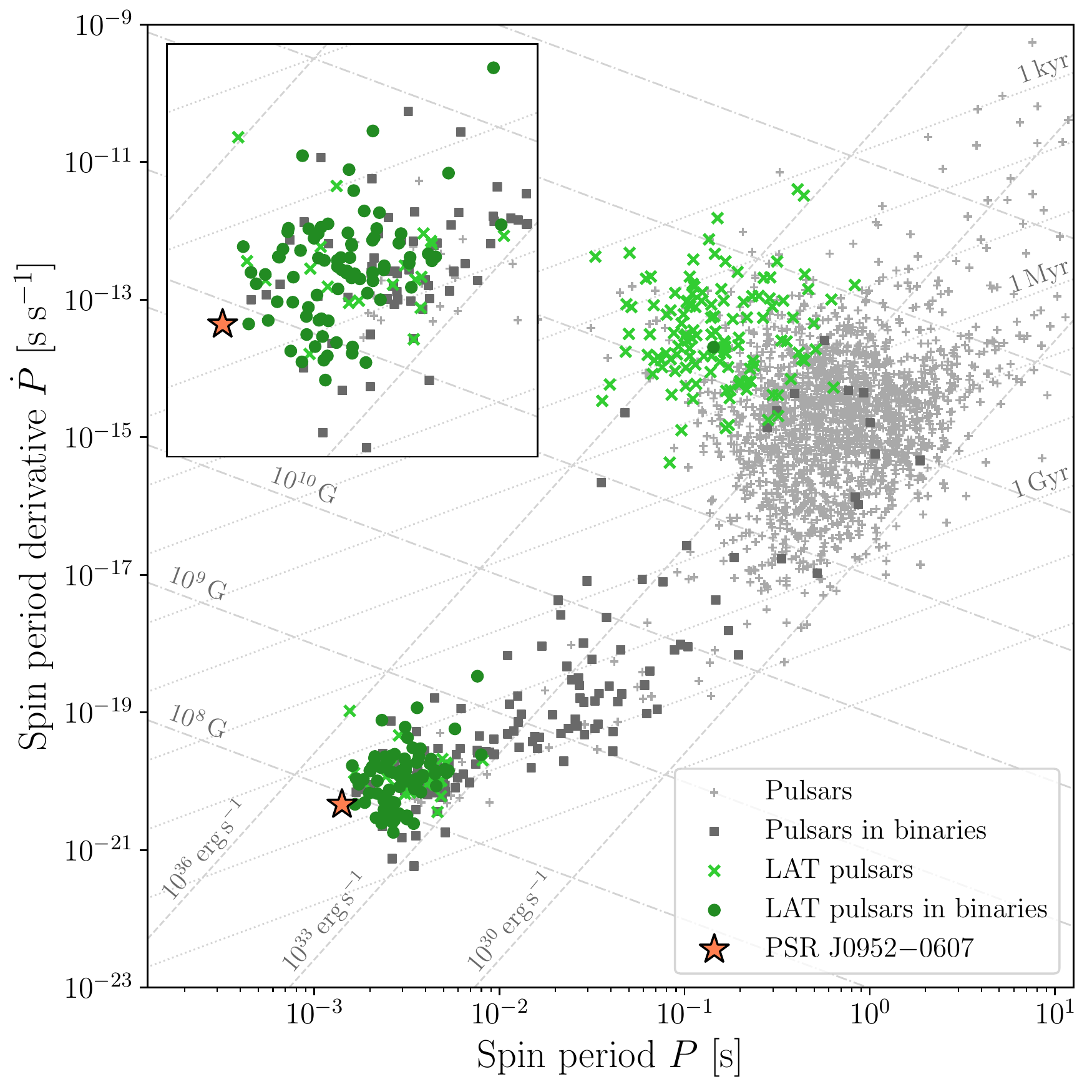}
			\hfill
		}
		\caption{\label{f:ppdot}
			Spin period $P$ and spin-period derivative $\dot{P}$ of the known pulsar population outside of globular clusters. The inset shows a zoomed-in view of the known \ac{MSP} population. Isolated radio pulsars (light-gray \replaced{crosses}{pluses}), binary radio pulsars (dark-gray \replaced{circles}{squares}), isolated gamma-ray pulsars (light-green crosses) and binary gamma-ray pulsars (dark-green circles) are shown. The subject of this paper, the gamma-ray pulsar \psr{}, is marked by an orange star. The lines denote constant characteristic age $\tau_\text{c}$ (dotted), spin-down power $\dot{E}$ (dashed) and surface magnetic field strength $\Bsurf$ (dashed-dotted).
		}
	\end{figure}

	In Figure~\ref{f:ppdot}, \psr{} is shown in a $P$-$\dot{P}$ diagram with the known pulsar population outside of globular clusters. The spin parameters of the more than $2000$ radio pulsars are taken from the ATNF Pulsar Catalogue\textsuperscript{\ref{n:atnf}} \citep{manchester2005}.
	
	Furthermore we estimated the characteristic age $\tau_\text{c}$, the spin-down power $\dot{E}$, the surface magnetic field strength $\Bsurf$ and the magnetic field strength at the light cylinder $B_\text{LC}$ (see Table~\ref{t:timing}). To calculate these values we assumed the pulsar to be a magnetic dipole with a canonical radius $r_\text{psr} = 10 \, \text{km}$ and moment of inertia $I_\text{psr} = 10^{45} \, \text{g} \, \text{cm}^2$ \citep[e.g.,][]{abdo2013}. The same assumptions were used to plot the contour lines in Figure~\ref{f:ppdot}.
	
	Despite spinning so rapidly, \added{the gamma-ray energy flux of} \psr{} is on the fainter end of the gamma-ray \ac{MSP} population. There are several reasons why gamma-ray pulsars might appear faint, including large distance, high background, or low luminosity \citep{hou2014}. \psr{} is not in a high-background region. The large distance derived from the optical modeling could be a possible explanation but disagrees with the distance estimates derived from the dispersion measure, $d = (0.97,1.74)\,\text{kpc}$ (\citetalias{NE2001}, \citetalias{YMW16}). The inferred gamma-ray luminosity is \replaced{$L_\gamma = 4 \pi d^2 F_{\gamma} f_\Omega \approx (2.9,9.4) \times 10^{32} \, \textrm{erg} \, \textrm{s}^{-1}$}{$L_\gamma = 4 \pi d^2 F_{\gamma} f_\Omega \approx 3.1 \times 10^{32} \times (d/1\,\textrm{kpc})^2 \, \textrm{erg} \, \textrm{s}^{-1}$}. The measured \ac{LAT} energy flux $F_{\gamma}$ is given in Table~\ref{t:spectralanalysis} and we assumed no beaming (i.e., $f_\Omega = 1$). \replaced{Thus for the dispersion-measure derived distances the gamma-ray efficiency of $\eta_\gamma = L_\gamma / \dot{E} \approx (0.5\%,1.5\%)$ would be at the lower end of the distribution of other \acp{MSP} \citep{abdo2013}.}{The gamma-ray efficiency is $\eta_\gamma = L_\gamma / \dot{E} \approx 0.5\% \times (d/1\,\textrm{kpc})^2$. At the optical distance, $\eta_\gamma \approx 16\%$ is typical of gamma-ray \acp{MSP} \citep{abdo2013}, while at the DM-derived distance, $\eta_\gamma \sim 1\%$ would be unusually low.}
	
	Due to the non-detection of \psr{} in X-rays \citep[$F_\text{X} < 1.1 \times 10^{-13} \, \text{erg} \, \text{s}^{-1} \, \text{cm}^{-2}$,][]{bassa2017b} we can only give a lower limit for the gamma-ray-to-X-ray flux ratio $F_{\gamma}/F_\text{X} > 20$. This limit is at the lower end of the observed distribution but still consistent with the literature \citep{marelli2011,marelli2015,abdo2013,salvetti2017}.
	
	The peak of the observed optical light curve is fairly broad in orbital phase. This requires either low inclination such that part of the heated face of the companion is visible over a large range of orbital phases, or for the companion to be close to filling its Roche lobe, such that the tidal deformation results in an ``ellipsoidal'' component peaking at $\phiorb = 0.5$ and $\phiorb = 1.0$ (with $\phiorb = 0$ corresponding to the pulsar's ascending node) where the visible surface area of the companion is largest. Our best-fitting \texttt{Icarus} model favors the latter explanation, with $f_{\rm RL} \approx 88$\% and $i \approx 61 \degr$. However, high filling factors imply a larger and hence more luminous companion, and therefore require greater distance, with our model having $d \sim 4.7$--$6.6$\,kpc. \deleted{This disagrees strongly with the DM-derived distance estimates of $d=0.97$\,kpc (\citetalias{NE2001}) and $d=1.74$\,kpc (\citetalias{YMW16}).}
	
	We tried to re-fit the optical light curve with the distance fixed at the \citetalias{YMW16} distance of $d=1.74$\,kpc, but the resulting model has a significantly worse fit, and the low filling factor required results in an extremely high volume-averaged density for the companion ($\rho$) in excess of $100$\,g\,cm$^{-3}$. For comparison, the densest known black-widow companions have densities of around $50$\,g\,cm$^{-3}$ \citep[e.g. PSR\,J0636$+$5128][]{kaplan2018}, with the record being that of the black-widow candidate 3FGL~J1653.6$-$0158 in a 75-min orbit \citep{Romani2014+J1653} where $\rho \gtrsim 70$\,g\,cm$^{-3}$. These objects have been proposed to be the descendants of ultra-compact X-ray binaries, but this origin is unlikely for \psr{} given its much longer orbital period \citep{vanHaaften2012+UCXBs}. If the DM distances are assumed, the required density suggests that the companion star consists mostly of degenerate matter. A low filling factor may also explain the absence of radio eclipses seen from \psr{}. Alternatively, the low-density, large-distance solution has $\rho \sim 2.75$\,g\,cm$^{-3}$, close to the density of brown dwarfs of similar mass and temperature given by the model considered in \citet{kaplan2018}.

	We note that similar discrepancies in model distances were seen by \citet{Sanchez2017+Bduct} when using a direct-heating model. \citet{Romani2016+ShockHeating} and \citet{Sanchez2017+Bduct} considered models that additionally include a contribution from reprocessing of the pulsar wind by an intra-binary shock, which can wrap around the companion star. This can produce broader light curves for lower filling factors as some heating flux is re-directed further around the sides of the companion star, and can also explain the small phase offset required for our direct-heating model by asymmetry in the shock front. Such a model may improve the fit for lower distances and filling factors, although an extremely high companion density would still be required to match the \citetalias{YMW16} distance. A likely explanation therefore could be that some heating flux is reprocessed by a shock, and the system has a moderate distance and filling factor, somewhat larger than required by the \citetalias{YMW16} value, but below those predicted by our direct-heating model. While more complex irradiation models \citep[e.g.,][]{Romani2016+ShockHeating} may be required to address this issue, a full investigation of alternative models is beyond the scope of this study.

	In both the small and large distance cases, we find that the nightside temperature of the companion is $T_{\rm n} \approx 3000 \pm 250$\,K at $95\%$ confidence. We also find a well-constrained irradiating temperature of $T_{\rm irr} = 6100 \pm 350$\,K, higher than that found from the single-band fit performed in \citet{bassa2017b}. This heating parameter can be compared to the total energy budget of the pulsar by calculating the ``efficiency'', $\varepsilon$, of conversion between spin-down power ($\dot{E}$) and heating flux \citep{Breton2013+4MSPs}
	\begin{equation}
		\varepsilon = \frac{4 \pi A^2 \sigma T_{\rm irr}^4}{\dot{E}}\,,
	\end{equation}
	with $\varepsilon \sim 20$\% being typical for black-widow systems. The efficiency is also shown in Figure~\ref{f:Icarus_MCMC}, calculated from $T_{\rm irr}$ and from the orbital separation ($A = x\, (1 + q)/\sin i$) at each point. We find that heating represents a larger fraction of the pulsar's total energy budget ($\varepsilon \sim 22$\% to $48$\% with $95$\% confidence) than the observed gamma-ray emission $\eta_\gamma \approx 0.5\% \times (d/1 {\rm kpc})^2$. This estimate assumes that the pulsar's heating flux is emitted isotropically. As pointed out by \citet{Draghis2018+HeatedComp}, some models of pulsar gamma-ray emission predict stronger beaming towards the pulsar's rotational equator, and an \ac{MSP}'s rotation should be aligned with the orbital plane as a result of the spin-up process. The actual gamma-ray luminosity directed towards the companion may therefore be higher than we observe. Our optical fits suggest a relatively face-on inclination (further evidenced by the lack of eclipses observed in radio observations, which often occur far outside the companion's Roche lobe), and so the comparative faintness of the pulsar's observed gamma-ray emission could be explained by the large viewing angle, and the fact that flux is preferentially emitted in the equatorial plane. A full modeling of the pulsar's phase-aligned radio and gamma-ray pulse profiles would provide an additional test of this scenario by estimating the viewing and magnetic inclination angles, and the relative beaming factors along our line of sight and in the equatorial plane. So far this is inhibited by the low significance of the gamma-ray light curve but with the continuing \ac{LAT} mission this might be possible with more gamma-ray data in the future.
	
	Alternatively, the difference between the heating flux and gamma-ray emission may suggest that another mechanism, e.g., the pulsar wind or intra-binary shock heating \citep{Romani2016+ShockHeating,Wadiasingh2017+Shock}, is responsible for heating the companion. Indeed, there is evidence for this being the case for the transitional PSR\,J1023$-$0038 where the optical heating is apparently unchanged between the \ac{MSP} and low-mass-X-ray-binary (LMXB) states \citep{Kennedy2018+J1023K2} despite a $5\times$ increase in the gamma-ray flux \citep{stappers2014}.

	As the optical counterpart to \psr{} is faint (peaking at $r^\prime \approx 22$), it will be difficult to improve upon this picture of the system. While it may be possible to improve upon the dayside temperature measurement with optical spectroscopy in the future, the companion is effectively undetectable at minimum ($r^\prime > 25.0$), precluding optical spectroscopic measurements of the companion's nightside temperature. We are also unable to constrain the mass of \psr{} using the optical data. Constraining the pulsar mass would require a precise measurement of the binary mass ratio, which can be obtained for black-widow systems by comparing the radial velocities of the pulsar and companion. Unfortunately, the optical counterpart of \psr{} is too faint ($r^\prime \sim 23$ at quadrature when the radial velocity is highest) for spectroscopic radial velocity measurements to be feasible even with 10\,m class telescopes.
	
	The gamma-ray source shows no significant variability as all flux measurements are consistent with the mean flux level. The calculated variability index also indicates a non-varying source. Here it is important to note that due to the low flux of the source the time bins had to be $750$ days long to keep statistical precision. Therefore the variability index was calculated from only five independent time bins. Variations on shorter timescales can also not be found this way.
	
\begin{figure}
		\centerline{
			\hfill
			\includegraphics[width=0.99\columnwidth]{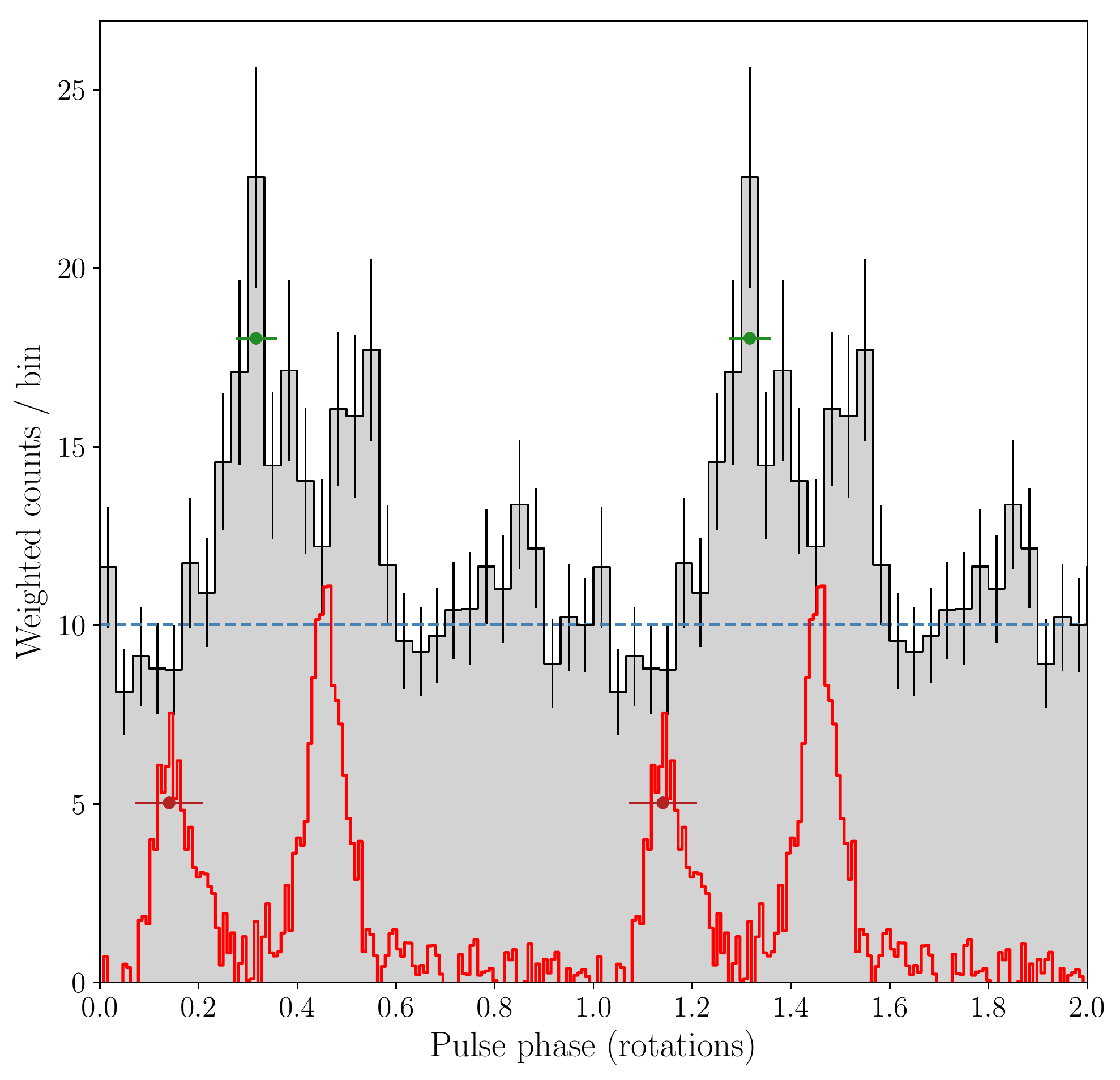}
			\hfill
		}
		\caption{\label{f:radio_align}
			Aligned integrated gamma-ray and radio pulse profiles of \psr{} over two identical rotations. The black curve shows the weighted \ac{LAT} photon counts after MJD\,$55{,}750$ in a histogram with $30$ bins per rotation. The green error bars show the phase uncertainty of the gamma-ray pulse profile. The estimated background level is indicated by the dashed blue line. The radio profile as seen by the \ac{LOFAR} telescope in a $78\,\textrm{MHz}$ band centered at $149\,\textrm{MHz}$ is drawn in red. The error bars drawn in dark red indicate the possible phase shift of the radio pulse profile due to a dispersion measure variation of $10^{-3}\,\textrm{pc}\,\textrm{cm}^{-3}$ over the time span of the \Fermi{} mission.
		}
	\end{figure}

	The gamma-ray pulse profile of \psr{} shows two peaks that are separated by $\mu_2 - \mu_1 \approx 0.2$ rotations. This is typical for gamma-ray \acp{MSP}. More than half of them are double peaked with a peak separation of $0.2$ -- $0.5$ rotations \citep{abdo2013}. The radio pulse profile also shows two peaks with similar separation, with the radio pulse slightly leading the gamma-ray pulse (see Figure~\ref{f:radio_align}). The phase lag between the gamma-ray and radio pulse profile seems to be $\sim 0.15$ \citep[the majority of two-peaked \acp{MSP} show phase lags of $0.1$ -- $0.3$;][]{abdo2013}. Due to a covariance between $f$ and dispersion measure (see Section~\ref{s:radio}) we were not able to measure significant variations in the dispersion measure. A change in dispersion measure of $10^{-3}\,\textrm{pc}\,\textrm{cm}^{-3}$ over the course of the \Fermi{} mission would lead to an error in the phase offset of $13\%$.

	\begin{figure}
		\centerline{
			\hfill
			\includegraphics[width=0.99\columnwidth]{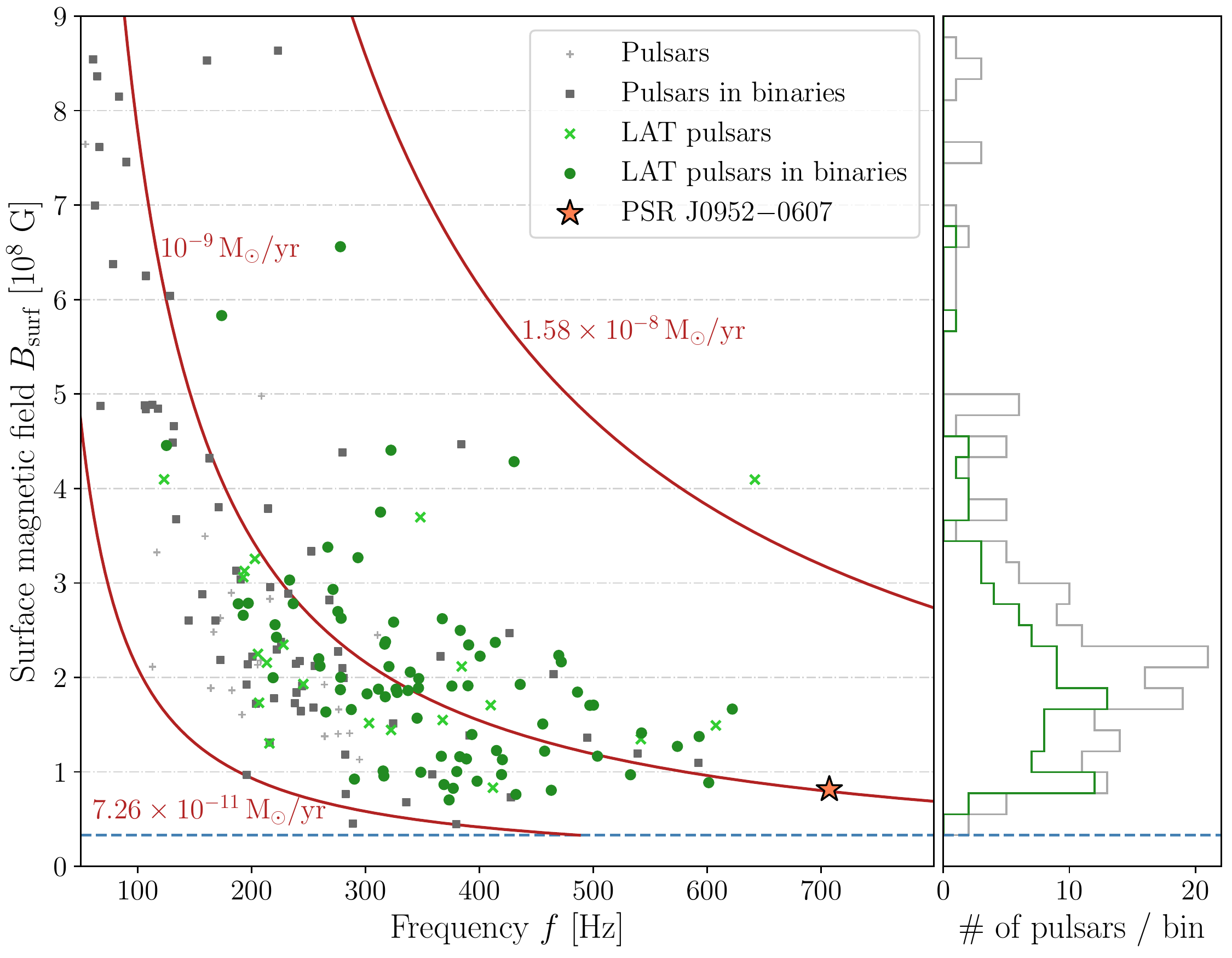}
			\hfill
		}
		\caption{\label{f:fbsurf}
			Frequency $f$ and surface magnetic field strength $\Bsurf$ of the known \ac{MSP} population outside of globular clusters. The surface magnetic field of \psr{} is computed assuming $\Pdot_\text{Shk} = 0$ and thus represents an upper limit. The horizontal \added{dashed} blue line represents a possible minimum magnetic field strength. The three red lines are so-called spin-up lines for different accretion rates. Left panel: The markers are defined as in Figure~\ref{f:ppdot}. Right panel: Histogram with $40$ bins between $3.3\times 10^{7}\,\textrm{G}$ and $9\times 10^{8}\,\textrm{G}$, showing the inferred surface magnetic field strengths for the known \ac{MSP} population (gray) and also the subset of \ac{LAT} pulsars (green).
		}
	\end{figure}

	Gamma-ray pulsars are a good way to identify the maximum spin frequency of neutron stars. Among the ten fastest Galactic field pulsars only one pulsar has not been detected in gamma rays. Until the discovery of the $707$ Hz pulsar \psr{}, the first \ac{MSP}, \origpsr{}, and the first black-widow pulsar, \fbwpsr{}, were the fastest-spinning gamma-ray pulsars known \citep{guillemot2012}. Still, the mass-shedding spin limit for neutron stars is typically placed much higher at around $1200\,\textrm{Hz}$ \citep{cook1994b,lattimer2004}. One mechanism that could prevent neutron stars from spinning up to higher frequencies is the emission of gravitational waves \citep[for a recent work on this subject see, e.g.,][]{gittins2018}. Another option could be that the spin-up torque might be smaller for faster pulsars with lower magnetic field strengths \citep{patruno2012,bonanno2015}.
	
	The estimated intrinsic spin-period derivative implies a very low surface magnetic field of $8.2\times10^7$\,G for \psr{}. Assuming non-zero proper motion would result in an even lower surface magnetic field estimate. Just nine pulsars, including the gamma-ray pulsar with the lowest surface magnetic field in the ATNF Pulsar Catalogue\textsuperscript{\ref{n:atnf}} \citep{manchester2005}, \lowBpsr{} \citep{bhattacharyya2013}, show lower inferred surface magnetic fields (Figure~\ref{f:fbsurf}). The surface $B$-field of the other recent \ac{LOFAR}-detected pulsar, \oLOFpsr{}, is only slightly stronger \citep{pleunis2017}. This might be a hint that pulsars with low $B$-fields also have steeper \added{radio} spectra.
	
	The pulsar distribution in Figure~\ref{f:fbsurf} indicates a lower limit on the magnetic field strength independent of the spin frequency. The equilibrium spin period as predicted by \cite{alpar1982} is $P_\textrm{eq} \propto B_\textrm{surf}^{6/7}\,R_\textrm{psr}^{18/7}\,M_\textrm{psr}^{-5/7}\,\dot{M}_\textrm{accr}^{-3/7}$ with pulsar radius $R_\textrm{psr}$, mass $M_\textrm{psr}$, and accretion rate $\dot{M}_\textrm{accr}$, which indicates that the lowest spin periods can be reached for low magnetic field strengths and high accretion rates. Nevertheless high accretion rates lead to a rapid decrease of the magnetic field strength and for low magnetic field strengths the angular momentum transfer is slower \citep{bonanno2015}. In order to spin up to millisecond periods a limiting magnetic field strength and accretion rate can be set as a result of the amount of time a neutron star can spend accreting matter being limited by the age of the universe \citep{pan2018}. For a neutron star with a mass of $1.4\,\textrm{M}_\odot$, a radius of $10\,\textrm{km}$ and a minimum accretion rate of $7.26\times10^{-11}\,\textrm{M}_\odot\,\textrm{yr}^{-1}$ we get a minimum magnetic field strength of $B_\textrm{surf} \gtrsim 3.3 \times 10^7\,\textrm{G}$, which is consistent with the observed pulsar population.
	
	No continuous gravitational waves are detected from \psr{}, which is to date the fastest-spinning pulsar targeted for gravitational wave emission. 
	The $95\%$ upper limit on the intrinsic gravitational wave amplitude is set to $\hUL = \hULval$. The corresponding upper limit on the ellipticity is $\epUL={\epULval \times (d/\distance) \times (10^{45}\,\textrm{g\,cm}^2/I)}$, where $I$ is the principal moment of inertia of the pulsar. The intrinsic gravitational wave amplitude at the detector needed to account for all of the spin-down energy lost due to gravitational wave emission is $\hSD = \hSDval \times (\distance/d) \times (I/10^{45}\,\textrm{g\,cm}^2)^{1/2}$, corresponding to an ellipticity of $\epSD = \epSDval \times (\distance/d)$.
	
	As for many other high-frequency pulsars, the indirect spin-down upper limit \added{on $h_0$} is smaller and \added{more constraining} than our measured gravitational wave upper limit, in this case \added{by} a factor of $\approx$ \aboveSpindownLimitFactor{} \added{at $1\,\textrm{kpc}$. For a more likely larger distance the factor would be even greater}, so it is not surprising that a signal was not detected \citep{LVC2019}. The quoted spin-down upper limit could be inaccurate if the measured spin down were affected by radial motions, if the distance were smaller than estimated or if the moment of inertia of the pulsar were different than the fiducial value of $10^{45}\,\textrm{g\,cm}^2$. In the case of \psr{} it is unlikely that all these effects could bridge a gap of nearly two orders of magnitude, but in line with the ``eyes-wide-open" spirit of previous searches for gravitational waves from known pulsars (see \cite{LVC2019,abbott2017knownpulsars, aasi2014} and references therein) we all the same perform the search.

	\section{Conclusions}\label{s:conclusions}
	
	Using a sensitive, fully coherent pulsation search technique, we detected gamma-ray pulsations from the radio pulsar \psr{} in a search around the parameters reported by \cite{bassa2017b}. New timing methods were developed to cope with the low signal strength, allowing us to measure the spin rate, sky position, and orbital period with high precision, and in agreement with the updated radio-timing ephemeris. Furthermore thanks to the longer gamma-ray time span we reliably constrained the intrinsic spin-period derivative $\Pdot_\text{int} \lesssim 4.6 \times 10^{-21} \, \text{s} \, \text{s}^{-1}$. This measurement provides estimates of physical parameters such as the spin-down luminosity ($\dot{E} \lesssim 6.4 \times10^{34}$\,erg s$^{-1}$), and a surface magnetic field ($\Bsurf \lesssim 8.2\times10^7$\,G) among the lowest of any detected gamma-ray pulsar. Although the resulting timing solution spans $7$ years to the present data, we were unable to extend this to cover data earlier than MJD\,$55{,}750$. We investigated several possible reasons. Flux variations could lead to the loss of pulsations. A time-varying orbital period as seen in several spider pulsars would cause a loss of phase coherence. With our current data we are not able to ascertain the true reason. In the absence of orbital-period variations or state changes, improved timing precision from additional data should help determine the cause.

	We also obtained new multi-band photometry of the pulsar's optical counterpart, and modeled the resulting light curve. To explain the observed optical flux, our models require either a much larger distance ($\sim5$\,kpc) than the DM-distance estimates of $0.97$\,kpc \citepalias{NE2001} to $1.74$\,kpc \citepalias{YMW16}, or a small and extremely dense companion $\rho \gg 100$\,g\,cm$^{-3}$. More complex optical models including intra-binary shocks might help to solve this discrepancy, but a full investigation of other models is beyond the scope of this work. We found that the pulsar flux heating the companion star accounts for a much larger fraction of the pulsar's spin-down power ($\sim 50$\%) than is converted to observed gamma-ray emission ($0.5$\% at $1$\,kpc), although this difference is reduced if our larger distance estimate is adopted.
	
	Despite the extensive analysis of \psr{} and its companion, the study of this pulsar has not ended as some questions remain unanswered. The \ac{LAT} and \ac{LOFAR} continue to take gamma-ray and radio data on this source, and we plan to obtain more optical data.
	
	\ac{LAT} gamma-ray data has helped to find many new \acp{MSP} by providing promising candidates \citep{ray2012}. Sophisticated methods to identify more pulsar candidates within \ac{LAT} sources have been developed \citep[e.g.,][]{Lee2012+GMM,parkinson2016}. For instance, \cite{frail2016a} identified $11$ promising \ac{MSP} candidates by checking for steep-spectrum radio sources coincident with \ac{LAT} sources. With the approach successfully used in this paper, new binary \ac{MSP} candidates can be searched for pulsations and upon detection the pulsar can be precisely timed within months after its discovery. Identifying more of the rapidly rotating spider pulsars will be helpful to study further the observed neutron star parameter limits like the maximum spin frequency and the minimum surface magnetic field strength.

	\acknowledgements
	
	\added{We thank the referee for pointing out the uncertainties in the DM/distance models, and suggesting the arguments given around Figure~\ref{f:dmdistance}.}
	This work was supported by the Max-Planck-Gesellschaft (MPG) and the ATLAS cluster computing team at AEI~Hannover. C.J.C., R.P.B., and D.M.-S. acknowledge support from the ERC under the European Union's Horizon 2020 research and innovation programme (grant agreement No. 715051; Spiders). C.G.B. and J.W.T.H. acknowledge support from the European Research Council (ERC) under the European Union's Seventh Framework Programme (FP7/2007-2013)/ERC grant agreement No. 337062 (DRAGNET; PI: Hessels). This work was supported by an STSM Grant from COST Action CA16214. M.R.K. is funded through a Newton International Fellowship provided by the Royal Society. Work at NRL is supported by NASA.
	
	The \textit{Fermi} LAT Collaboration acknowledges generous ongoing support from a number of agencies and institutes that have supported both the development and the operation of the LAT as well as scientific data analysis. These include the National Aeronautics and Space Administration and the Department of Energy in the United States, the Commissariat \`a l'Energie Atomique and the Centre National de la Recherche Scientifique/Institut National de Physique Nucl\'eaire et de Physique des Particules in France, the Agenzia Spaziale Italiana and the Istituto Nazionale di Fisica Nucleare in Italy, the Ministry of Education, Culture, Sports, Science and Technology (MEXT), High Energy Accelerator Research Organization (KEK) and Japan Aerospace Exploration Agency (JAXA) in Japan, and the K.~A.~Wallenberg Foundation, the Swedish Research Council and the Swedish National Space Board in Sweden. This work performed in part under DOE Contract DE-AC02-76SF00515.
	
	Additional support for science analysis during the operations phase is gratefully acknowledged from the Istituto Nazionale di Astrofisica in Italy and the Centre National d'\'Etudes Spatiales in France.
	
	Part of this work is based on data obtained with the international LOFAR Telescope (ILT) under project codes LC7\_018, DDT7\_002, LT5\_003, LC9\_041, and LT10\_004. LOFAR (van Haarlem et al., 2013) is the Low Frequency Array designed and constructed by ASTRON. It has observing, data processing, and data storage facilities in several countries, that are owned by various parties (each with their own funding sources) and that are collectively operated by the ILT foundation under a joint scientific policy. The ILT resources have benefited from the following recent major funding sources: CNRS-INSU, Observatoire de Paris and Universit\'e d'Orl\'eans, France; BMBF, MIWF-NRW, MPG, Germany; Science Foundation Ireland (SFI), Department of Business, Enterprise and Innovation (DBEI), Ireland; NWO, The Netherlands; The Science and Technology Facilities Council, UK.
	
	HiPERCAM and V.S.D. are funded by the European Research Council under the European Union's Seventh Framework Programme (FP/2007-2013) under ERC-2013-ADG grant agreement number 340040 (HiPERCAM). ULTRACAM and V.S.D. are funded by the UK Science and Technology Facilities Council. This work is based on observations made with the Gran Telescopio Canarias (GTC), installed in the Spanish Observatorio del Roque de los Muchachos of the Instituto de Astrof\'{i}sica de Canarias, on the island of La Palma, and on observations made with ESO Telescopes at the La Silla Paranal Observatory.
	
	This research has made use of data, software and/or web tools obtained from the Gravitational Wave Open Science Center (https://www.gw-openscience.org), a service of LIGO Laboratory, the LIGO Scientific Collaboration and the Virgo Collaboration. LIGO is funded by the U.S. National Science Foundation. Virgo is funded by the French Centre National de Recherche Scientifique (CNRS), the Italian Istituto Nazionale della Fisica Nucleare (INFN) and the Dutch Nikhef, with contributions by Polish and Hungarian institutes.
	
	\added{\software{\textit{Fermi} Science Tools, \textsc{dspsr} \citep{straten2011}, \textsc{psrchive} \citep{hotan2004}, \textsc{tempo2} \citep{hobbs2006,edwards2006}, NE2001 \citep{NE2001}, YMW16 \citep{YMW16}, \texttt{MultiNest} \citep{MultiNest}, \texttt{PyMultiNest} \citep{PyMultiNest}, ULTRACAM/HiPERCAM software pipelines, \texttt{Icarus} \citep{Breton2012+Icarus}, \texttt{psrqpy} \citep{manchester2005,pitkin2018}, \texttt{Astropy} \citep{astropy2013,astropy2018}, \texttt{matplotlib} \citep{matplotlib2007}, \texttt{NumPy} \citep{numpy2006,numpy2011}}}
	
	\appendix
	\section{Estimating the false-alarm probability for a multi-dimensional $H$ statistic search}\label{a:signif}

	It is important to estimate the false-alarm probability $P_\text{FA}$ to know if the gamma-ray detection is real. As described in Section~\ref{s:detection}, there is no known analytical expression for the false-alarm probability of the maximum value from an $H$ statistic search over a dense, multi-dimensional parameter grid. Deriving the probability distribution for the maximum value of a multi-dimensional ``random field'' is difficult and approximate solutions are only known for simple cases such as Gaussian or chi-squared random fields \citep{RFG}. While the power in a single harmonic does follow a chi-squared random field in the presence of random noise, the known solutions cannot be applied in this case due to the maximization over summed harmonics and penalty factors defining the $H$ statistic, and the fact that the metric density varies between different summed harmonics. Even for chi-squared random fields, there is no simple ``trials factor'' that can be applied to the single-trial false-alarm probability (which for the $H$ statistic was derived by \citealt{kerr2011}): the false-alarm probability depends on the volume, shape, and dimensionality of the search space \citep{RFG}. A full discussion of this is beyond the scope of this work.  Below, we show empirically that a simple trials factor approach over-estimates the detection significance, and describe the ``bootstrapping'' method that we used to overcome this.
  
	The false-alarm probability for a single $H$ statistic trial is
	\begin{equation}
		P_\text{FA}(H_\text{m}\,|\,a) = \mathrm{e}^{-a\, H_\text{m}},
		\label{eq:pfa_st}
	\end{equation}
	with scaling factor $a = 0.3984$ \citep{dejager2010,kerr2011}. This formula can be used to estimate the significance of the maximum $H$ statistic value after $n$ \textit{independent} trials
	\begin{equation}
		P_\text{FA}(H_\text{m}\,|\,a,n) = 1 - \left[1 - \mathrm{e}^{-a\, H_\text{m}}\right]^n \,.
		\label{eq:pfa_tot}
	\end{equation}

	We assume at first that our search contained a number of ``effective'' independent trials ($N_{\rm eff}$) that is some unknown fraction of the number of actual trials (i.e. the number of grid points at which we evaluated the $H$ statistic). We then estimated $N_{\rm eff}$ from the results of our search as follows. We divided our parameter space into $n_{\rm seg} =2 \times 17 \times 13 = 442$ segments in $f$, $\fdot$, and $\Porb$ respectively. The number of segments in $f$ and $\fdot$ is determined by the parameter space volumes, which were searched in parallel, as only the highest $H$ statistic values from each were stored. To ensure that all segments were independent from the pulsar signal, we removed all grid points within those segments which were close (according to the parameter space metric; see Section~\ref{s:search}) to the pulsar parameters.

\begin{figure}
		\centerline{
			\hfill
			\includegraphics[width=0.5\columnwidth]{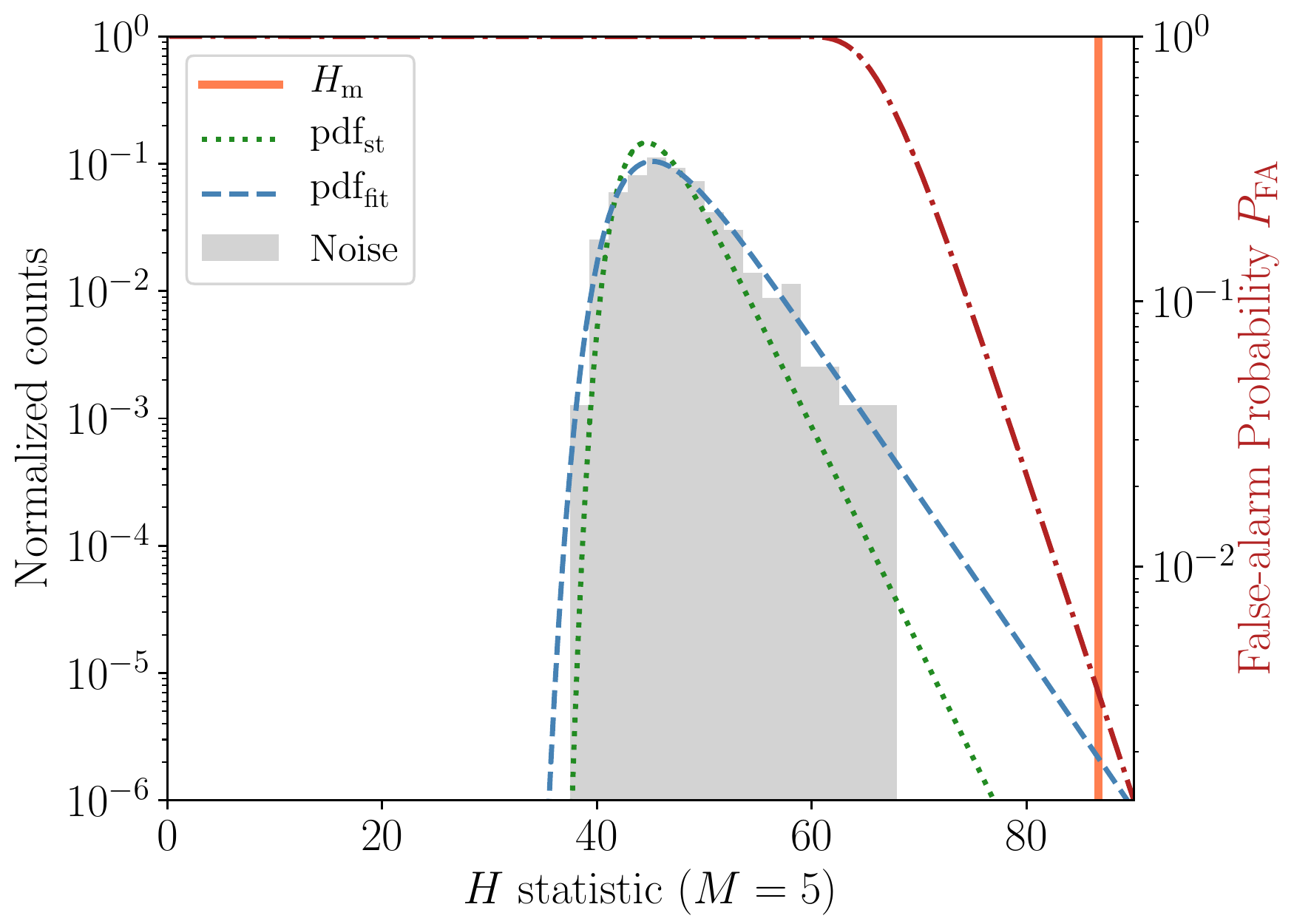}
			\hfill
		}
		\caption{\label{f:signif}
		Normalized histogram showing the highest $H$ statistics for $442$ subsets of our search space after excluding results affected by the pulsar signal. The dotted green and dashed blue curves show normalized probability density functions for the maximum $H$ statistic obtained after $n$ effective trials. The curves gave maximum likelihood after varying over $n$ with fixed single-trial scaling factor $a=0.3984$ (dotted green) and after varying $a$ and $n$ jointly (dashed blue). The maximum $H$ statistic for the pulsar $H_\text{m} = 86.7$ is marked by the vertical orange line. The red line (dashed-dotted) shows the false-alarm probability depending on $H_\text{m}$ computed with Equation~\eqref{eq:pfa_tot} with $a$ and $n$ from the joint variation.
		}
	\end{figure}

	The highest $H$ statistic of each of the segments is plotted in the normalized histogram in Figure~\ref{f:signif}. We fit for the effective number of trials \citep[as done by, e.g.,][]{kruger2002} by maximizing the likelihood, 
   	\begin{equation}
     		L(n,a|H_{m,i}) = \prod_{i} p(H_{m,i} | a,n)\,
	\end{equation}
    for our set of $H$ statistic values, according to the probability density function for $H_\text{m}$ after $n$ trials (the derivative of Equation~\eqref{eq:pfa_tot}),
	\begin{equation}
		p(H_\text{m}|a,n) = a \, n  \left[ 1 - \mathrm{e}^{-a\,H_\text{m}} \right]^{n-1} \, \exp(-a\,H_\text{m}) \,.
	\end{equation}
	However, as shown in Figure~\ref{f:signif}, the tail of the best fitting distribution is significantly under-estimated, leading to \textit{over-estimated} significances for large $H$ statistic values. This demonstrates that there is no simple effective trials factor that can be applied to estimate the overall significance.

	To overcome this, we performed a second fit, maximizing over the likelihood for both $n$ and $a$. The resulting best-fitting distribution is also shown in Figure~\ref{f:signif}. We found the best-fitting scaling factor to be $\hat{a} \approx 0.284$, meaning the probability density function is flatter and gives a more conservative estimate for the significance. We note that this should not apply in general, and will depend, amongst other factors, on the dimensionality of the search space and the number of harmonics summed. 

	Finally, we use $\hat{a}$ and multiply the best-fitting $n$ (the best-fitting per-segment trials factor) by $n_{\rm seg}$, and apply Equation~\eqref{eq:pfa_tot} to obtain an approximation to the false-alarm probability for the maximum $H$ statistic value. For the candidate pulsar signal, this was $P_\text{FA} = 0.33\%$. For comparison the candidate with the largest $H$ statistic from a segment of the search not affected by the pulsar signal had $P_\text{FA} = 56\%$.

	\bibliographystyle{aasjournal}
	
	\bibliography{library}
	
	\listofchanges
	
\end{document}